\documentclass[prb,reprint,amssymb]{revtex4-2}

\usepackage{xcolor}

\usepackage{bm}

\usepackage{amsmath}  
\usepackage{amsfonts}
\usepackage{graphicx}
\usepackage{hyperref}
\hypersetup{
    colorlinks=true,
	citecolor=blue,
    linkcolor=red,
	urlcolor = blue
}

\begin{document}

\title{Nonlocal damping of spin waves in a magnetic insulator induced by 
normal, heavy, or altermagnetic metallic overlayer: A Schwinger-Keldysh field theory approach}

\author{Felipe Reyes-Osorio}
\author{Branislav K. Nikoli\'c}
\email{bnikolic@udel.edu}
\affiliation{Department of Physics and Astronomy, University of Delaware, Newark, DE 19716, USA}


\begin{abstract}
Understanding spin wave (SW) damping, and how to control it to the point of being able to amplify SW-mediated signals, is one of the key requirements to bring the envisaged magnonic technologies to fruition. Even widely used magnetic insulators with low magnetization damping in their bulk, such as yttrium iron garnet, exhibit \textit{100-fold increase} in SW damping due to inevitable contact with metallic layers in magnonic circuits, as observed in very recent experiments [I. Bertelli \textit{et al.}, Adv. Quantum Technol. \textbf{4}, 2100094 (2021)] mapping SW damping in spatially-resolved fashion. Here, we provide microscopic and rigorous understanding of \textit{wavevector-dependent} SW damping using extended Landau-Lifshitz-Gilbert equation with \textit{nonlocal damping tensor}, instead of conventional local scalar Gilbert damping, as derived from Schwinger-Keldysh nonequilibrium quantum field theory. In this picture, the origin of nonlocal magnetization damping and thereby induced wavevector-dependent SW damping is interaction of localized magnetic moments of magnetic insulator with conduction electrons from the examined three different types of metallic overlayers---normal, heavy, and altermagnetic. Due to spin-split energy-momentum dispersion of conduction electrons in the latter two cases, the nonlocal damping is anisotropic in spin and space, and it can be dramatically reduced by changing the relative orientation of the two layers when compared to the usage of normal metal overlayer.
\end{abstract}

\maketitle

\section{Introduction}

Spin wave (SW) or magnon damping is a problem of great interest to both basic and applied research. For basic research, its measurements~\cite{Li2016,Chen2022,Dai2000, Bayrakci2013} can reveal microscopic details of boson-boson or boson-fermion quasiparticle interactions in solids, such as:  magnon-magnon interactions (as described by second-quantized Hamiltonians containing products of three or more bosonic operators~\cite{Zhitomirsky2013,Bajpai2021}), which are frequently encountered in antiferromagnets~\cite{Bayrakci2013, Zhitomirsky2013} and quantum spin liquids~\cite{Smit2020, Winter2017}, wherein they play a much more important role~\cite{Gohlke2023} than boson-boson interactions in other condensed phases, like anharmonic crystalline lattices or superfluids~\cite{Zhitomirsky2013}; magnon-phonon interactions~\cite{Dai2000}, especially relevant for recently discovered two-dimensional magnetic materials~\cite{Chen2022}; and magnon-electron interactions in magnetic metals~\cite{Li2016, Hankiewicz2008, Tserkovnyak2009,Nikolic2021,Isoda1990,Buczek2009}.  For the envisaged magnon-based digital and analog computing technologies~\cite{Chumak2015,Csaba2017,Chumak2019,Chumak2022,Mahmoud2020}, understanding magnon damping makes it possible to develop schemes to  suppress~\cite{Hamadeh2014} it, and, furthermore, achieve amplification of nonequilibrium fluxes of magnons~\cite{Akhiezer1964,Evelt2016,Demidov2020,Breitbach2023}. In fact, overcoming damping and 
achieving amplification is the {\em key} to enable complex magnon  circuits where, e.g., a logic gate output must be able to drive the input of multiple follow-up gates. Let us recall that the concept of SW was introduced by Bloch~\cite{Bloch1930} as a wave-like disturbance in the local magnetic ordering of a magnetic material. The quanta~\cite{Bajpai2021} of energy of SWs of frequency $\omega$ behave as quasiparticles termed magnons, each of which carries energy $\hbar \omega$ and spin $\hbar$. As regards terminology, we note that in  magnonics~\cite{Chumak2015} SW is often used for excitations driven by antennas~\cite{Bertelli2021,Mae2022,Krysztofik2022, Serha2022} and/or described by the classical Landau-Lifshitz-Gilbert (LLG) equation~\cite{Evans2014,Kim2010,Hankiewicz2008,Tserkovnyak2009}, whereas magnon is used for the quantized version of the same excitation~\cite{Zhitomirsky2013}, or these two terms are used interchangingly.

\begin{figure}
    \centering
    \includegraphics[width=\columnwidth]{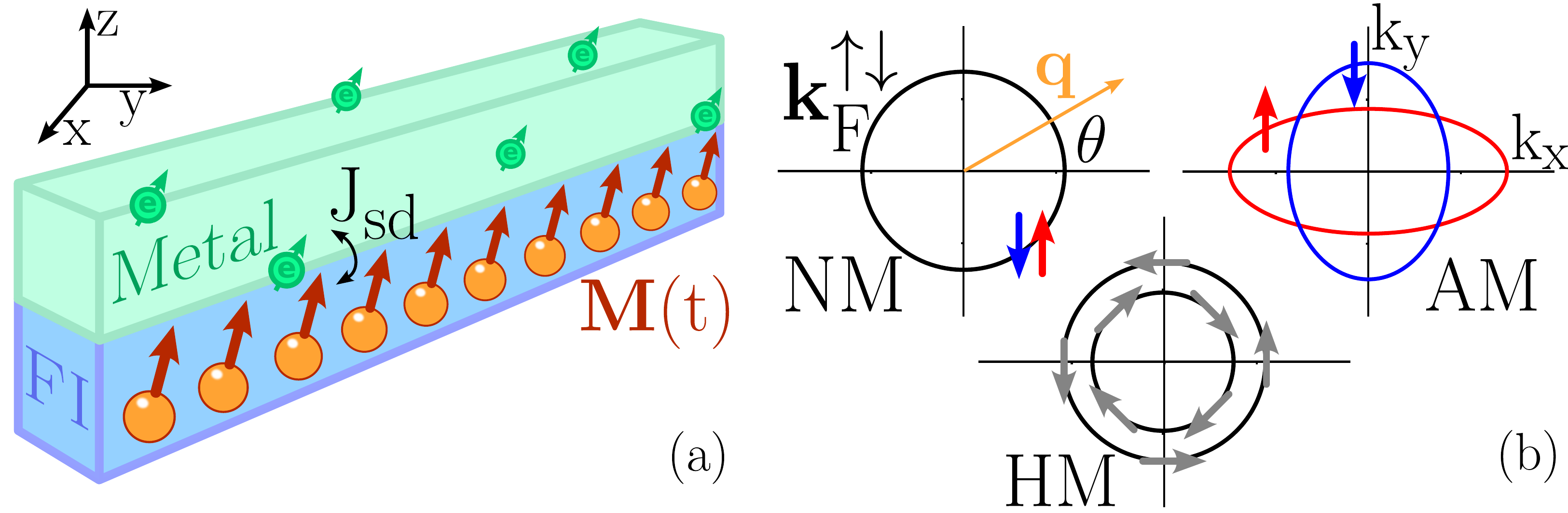}
    \caption{(a) Schematic view of bilayers where a metallic overlayer covers the top surface of magnetic insulator, as often encountered in spintronics and magnonics~\cite{Chumak2015, Serga2010}. Three different energy-momentum dispersion of conduction electrons at the interface are considered, with their Fermi surfaces shown in panel (b)---normal metal (NM); heavy metal (HM) with the Rashba SOC~\cite{Bihlmayer2022, Manchon2015}, and altermagnetic metal (AM)~\cite{Smejkal2022b, Smejkal2022a}---with the latter two being spin-split. The relative alignment of the layers is labeled by an angle $\theta$~\cite{Smejkal2022b, Smejkal2022a}, meaning that the wavevector $\mathbf{q}$ of  SWs within FI is at an angle $\theta$ away from the $k_x$-axis.}
    \label{fig:setup}
\end{figure}

In particular, experiments focused on SW damping in metallic ferromagnets have observed~\cite{Li2016} its dependence on the wavevector $\mathbf{q}$ which cannot be explained by using the standard LLG equation~\cite{Evans2014,Kim2010}, \mbox{$\partial_t \mathbf{M}_{n} = -\mathbf{M}_{n} \times \mathbf{B}^{{\rm eff}}_{n}  + \alpha_G \mathbf{M}_{n} \times \partial_t \mathbf{M}_{n}$} (where $\partial_t \equiv \partial/\partial t$), describing dynamics of localized magnetic moments (LMMs) $\mathbf{M}_n$ at site $n$ of crystalline lattice (also used in atomistic spin dynamics~\cite{Evans2014}) viewed as classical vectors of unit length. This is because $\alpha_G$, as the Gilbert damping parameter~\cite{Gilbert2004,Saslow2009}, is a \textit{local scalar} (i.e., position-independent constant). Instead, various forms of 
spatially {\em nonuniform} (i.e., coordinate-dependent) and {\em nonlocal} (i.e., magnetization-texture-dependent) damping due to conduction electrons have been proposed~\cite{Hankiewicz2008,Tserkovnyak2009,Zhang2009,Yuan2016,Verba2018}, or extracted from first-principles calculations~\cite{Lu2023}, to account for observed wavevector-dependent damping of SWs, such as 
$\propto q^2$ ($q=|\mathbf{q}|$) measured in Ref.~\cite{Li2016}. The  nonlocal damping terms require neither spin-orbit coupling (SOC) nor magnetic disorder scattering, in contrast to $\alpha_G$ which is considered to vanish~\cite{Gilmore2007} in their absence. 

Thus, in magnonics, it has been considered~\cite{Serga2010} that usage of magnetic  insulators, such as yttrium iron garnet (YIG) exhibiting  ultralow  $\alpha_G \simeq 10^{-4}$ (achieved on a proper substrate~\cite{Trempler2020}), is critical to evade much larger and/or nonlocal damping of SWs found in ferromagnetic metals. However,  very recent experiments~\cite{Bertelli2021,Mae2022,Krysztofik2022, Serha2022} have observed  {\em 100-fold} increase of SW damping in the segment of YIG thin film that was covered by a metallic overlayer. Such spatially-resolved measurement~\cite{Bertelli2021} of SW damping was made possible by the advent of quantum sensing  based on nitrogen vacancy (NV) centers in diamond~\cite{Casola2018}, and it was also subsequently confirmed by other methods~\cite{Mae2022,Krysztofik2022, Serha2022}. Since excitation, control, and detection of SWs requires to couple YIG to metallic electrodes~\cite{Chumak2015}, understanding the origin and means to 
control/suppress large increase in SW damping underneath metallic overlayer is crucial for realizing  magnonic technologies.

To explain their experiments, Refs.~\cite{Bertelli2021,Mae2022,Krysztofik2022, Serha2022} have  employed the LLG equation  with {\em ad hoc} introduced terms (such as, effective magnetic field due to SW induced eddy currents within metallic overlayer~\cite{Bertelli2021}) that can fit their experimental data. This approach is nonuniversal and unsatisfactory (many other examples of similar phenomenological strategy exist~\cite{Krivorotov2007,Li2016}). For example, simple renormalization of $\alpha_G$, as attempted in Ref.~\cite{Bertelli2021}, \textit{cannot}~\cite{Verba2018} capture properly dependence~\cite{Li2016, Bertelli2021, Tserkovnyak2009} of SW damping on wavevector, while postulating forms of spatially-dependent nonlocal damping~\cite{Hankiewicz2008,Tserkovnyak2009,Zhang2009,Yuan2016,Verba2018} leads to many ambiguous choices~\cite{Yuan2016}. A more microscopic route was taken in Refs.~\cite{Kapelrud2013, Skarsvaag2014} using the picture of spin angular momentum loss via spin pumping, but they predict only modest $\lesssim 2$-fold increase of SW damping (see Appendix~\ref{app:a}) and its independence on wavevector at small $q$, thereby \textit{contradicting 100-fold} increase found experimentally~\cite{Bertelli2021} or sensitive dependence on small wavevector values.

In contrast, in this paper we employ recently derived 
\begin{equation}\label{eq:LLGmodified}
    \partial_t \mathbf{M}_{n} = -\mathbf{M}_{n} \times \mathbf{B}^{{\rm eff}}_{n}  + \mathbf{M}_{n} \times \sum_{n^\prime} (\alpha_G\delta_{nn^\prime} + \lambda_{\mathbf{R}})\cdot\partial_t \mathbf{M}_{n^\prime},
\end{equation}
{\em extended} LLG equation with all terms obtained~\cite{ReyesOsorio2023a} microscopically from Schwinger-Keldysh nonequilibrium quantum field theory~\cite{Kamenev2011} and confirmed~\cite{ReyesOsorio2023a} via exact quantum-classical numerics~\cite{Petrovic2018,Bajpai2019,Petrovic2021,Suresh2020}. It includes nonlocal damping as the third term on the right-hand side (RHS), where its nonlocality is signified by dependence on $\mathbf{R}=\mathbf{r}_n-\mathbf{r}_{n^\prime}$, where $\mathbf{r}_n$ is the position vector of lattice site $n$. The Schwinger-Keldysh field theory (SKFT), commonly used in high energy physics and cosmology~\cite{Haehl2017, Haehl2017a, Berges2015}, allows one to ``integrate out" unobserved degrees of freedom, such as the conduction electrons in the setup of Fig.~\ref{fig:setup}, leaving behind a time-retarded dissipation kernel~\cite{Bajpai2019, Anders2022, Leiva2023} that encompasses electronic effects on the remaining degrees of freedom. This approach then rigorously  yields the effective equation for \textit{LMMs only}, such as Eq.~\eqref{eq:LLGmodified}~\cite{ReyesOsorio2023a, Leiva2023} which bypasses the need for adding~\cite{Li2016,Krivorotov2007,Bertelli2021} phenomenological wavevector-dependent terms into the standard LLG equation. In our approach, the nonlocal damping is extracted from the time-retarded dissipation kernel~\cite{ReyesOsorio2023a}.

This paper is organized as follows. Equation~\eqref{eq:LLGmodified} is applied to a setup depicted in Fig.~\ref{fig:setup} where conduction electron spins from three different choices for metallic overlayer are assumed to interact with LMMs of ferromagnetic insulator (FI) at the interface via $sd$ exchange interaction of strength $J_{sd}$, as well as possibly underneath the top surface of FI because of electronic evanescent wavefunction penetrating into it. Note that FI/normal metal (NM) bilayer directly models recent experiments~\cite{Bertelli2021} where FI was a thin film of YIG and NM was Au, and SW damping within FI was quantified using quantum magnetometry via NV centers in diamond. Next, the FI/heavy metal (HM) bilayer, such as YIG/Pt~\cite{Serha2022,Hamadeh2014}, is frequently encountered in various spintronics and magnonics phenomena~\cite{Chumak2015, Serga2010}. Finally, due to recent explosion of interest in altermagnets~\cite{Smejkal2022b, Smejkal2022a}, the FI/altermagnetic metal (AM) bilayers, such as YIG/RuO$_2$, have been explored experimentally to characterize RuO$_2$ as a spin-to-charge conversion medium~\cite{Bai2023}. 

\begin{figure}
    \centering
    \includegraphics[width=\linewidth]{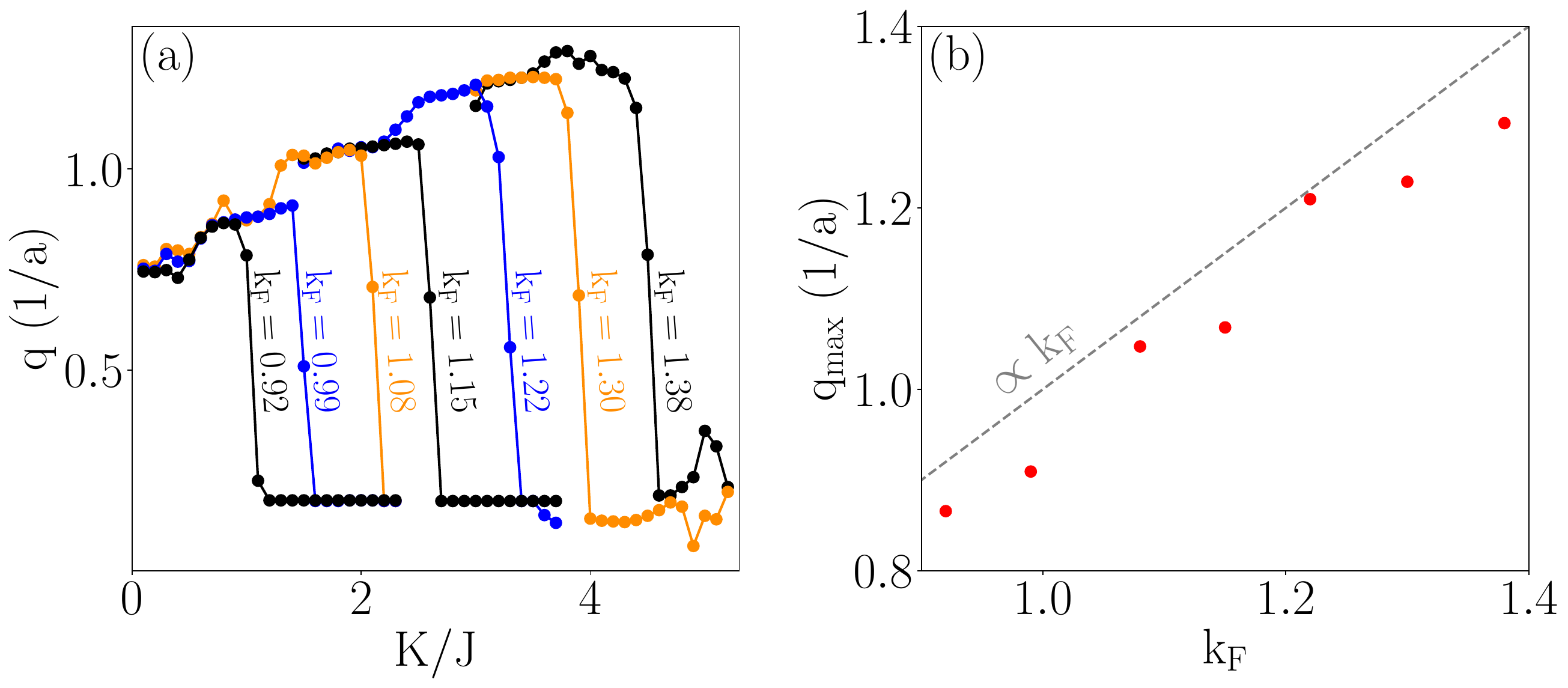}
    \caption{(a) Wavevector $q$ of SW generated by injecting spin-polarized current in TDNEGF+LLG simulations of NM overlayer on the top of 1D FI [Fig.~\ref{fig:setup}(a)] as a function of anisotropy $K$ [Eq.~\eqref{eq:energyFunctional}] for different electronic Fermi wavevectors $k_F$. (b) Maximum wavevector $q_{\rm max}$ of SWs that can be generated by current injection~\cite{Madami2011, Demidov2020} before wavevector-dependent SW damping becomes operative, as signified by the drop around $k_F$ in curves plotted in panel (a).}
    \label{fig:1Ddamping}
\end{figure}

\begin{figure*}

\includegraphics[width=\textwidth]{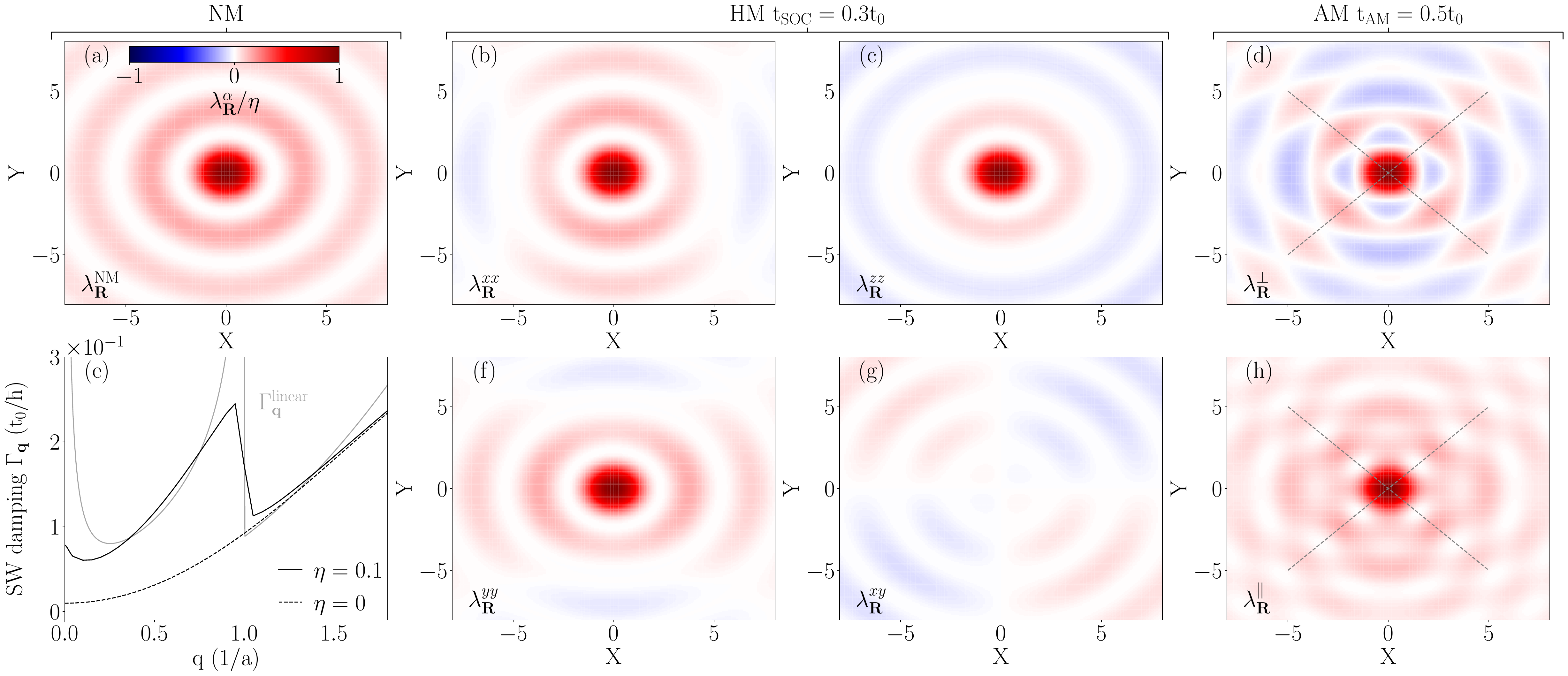}
    \caption{(a)--(d) and (f)--(h) Elements of SKFT-derived nonlocal damping tensor in 2D FI, $\lambda_\mathbf{R}$ where $\mathbf{R}=(X,Y,Z)$ is the relative vector between two sites within FI, covered by NM [Eq.~\eqref{eq:damping2dSpace}], HM [Eqs.~\eqref{eq:HMdamping}] or AM [Eqs.~\eqref{eq:AMdamp}] metallic overlayer. (e) Wavevector-dependent damping $\Gamma_\mathbf{q}$ of SWs due to NM overlayer, where the gray line is based on Eq.~\eqref{eq:damping2dWave} in the continuous limit~\cite{Halperin1969} and the other two lines are numerical solutions of extended LLG Eq.~\eqref{eq:LLGmodified} for discrete lattices of LMMs within FI. The dotted line in (e) is obtained in the absence of nonlocal damping ($\eta=0$), which is flat at small $q$.}
    \label{fig:2Ddamping}
\end{figure*}

\section{SKFT-based theory of SW damping in FI/metal bilayers}

The nonlocal damping~\cite{ReyesOsorio2023a} $\lambda_\mathbf{R}$ in the third term on the RHS of extended LLG Eq.~\eqref{eq:LLGmodified} stems from backaction of conduction electrons responding nonadiabatically~\cite{Bajpai2019,Sayad2015}---i.e., with electronic spin expectation value $\langle \hat{\mathbf{s}}_n \rangle$ being always somewhat behind LMM which generates spin torque~\cite{Ralph2008} $\propto \langle \hat{\mathbf{s}}_n \rangle \times \mathbf{M}_n$---to dynamics of LMMs. It is, in general, a nearly symmetric $3\times 3$ tensor whose components are given by~\cite{ReyesOsorio2023a}
\begin{equation}\label{eq:nonlocalDamping}
    \lambda^{\alpha\beta}_{\mathbf{R}} =  - \frac{J_{sd}^2}{2\pi} \int\!d\varepsilon \, \frac{\partial f}{\partial \varepsilon} \mathrm{Tr} \left [\sigma^\alpha A_{nn^\prime} \sigma^\beta A_{n^\prime n}\right].
\end{equation}
Here, $f(\varepsilon)$ is the Fermi function; $\alpha,\beta=x,y,z$; $\sigma^\alpha$ is the Pauli matrix; and $A(\varepsilon) = i\big[ G^R(\varepsilon) - G^A(\varepsilon)\big]$ is the spectral function in the position representation obtained from the retarded/advanced Green's functions (GFs) \mbox{$G^{R/A}(\varepsilon)=\big(\varepsilon - H \pm i\eta \big)^{-1}$}. Thus, the calculation of $\lambda_\mathbf{R}$ requires only an electronic Hamiltonian $H$ as input, which makes theory fully microscopic (i.e., Hamiltonian-based). In order to arrive at closed expressions for the nonlocal damping tensor, we employ simple model Hamiltonians, nevertheless, realistic materials can be treated by using the Hamiltonian input from first-principles calculations (as explained in Ref.~\cite{ReyesOsorio2023a}). Although the SKFT-based derivation~\cite{ReyesOsorio2023a} yields an additional antisymmetric term, not displayed in Eq.~\eqref{eq:nonlocalDamping}, such term vanishes if the system has inversion symmetry. Even when this symmetry is broken, like in the presence of SOC, the antisymmetric component is often orders of magnitude smaller~\cite{Leiva2023}, therefore, we neglect it. The first term on the RHS of extended LLG Eq.~\eqref{eq:LLGmodified} is the usual one~\cite{Evans2014,Kim2010}, describing precession of LMMs in the effective magnetic field, $\mathbf{B}^{{\rm eff}}_{n}$, which is the sum of both internal and external ($B_{\rm ext}\mathbf{e}_z$) fields. It is obtained as $\mathbf{B}^{\rm eff}_{n}=-\partial\mathcal H/\partial \mathbf{M}_{n}$ where $\mathcal{H}$ is the classical Hamiltonian of LMMs
\begin{equation}\label{eq:energyFunctional}
    \mathcal{H} = -J\sum_{\langle {nn}^\prime\rangle} \mathbf{M}_{n} \cdot \mathbf{M}_{n^\prime} + \frac{K}{2} \sum_{n} (M^z_{n})^2 - B_{\rm ext}\sum_{n} M^z_{n}.
\end{equation}
Here we use $g=1$ for gyromagnetic ratio, which simplifies Eq.~\eqref{eq:LLGmodified}; $J$ is the Heisenberg exchange coupling between the nearest-neighbors (NN) sites; and $K$ is the magnetic anisotropy.

When nonlocal damping tensor, $\lambda_\mathbf{R}$ is proportional to $3\times 3$ identity matrix, $\mathcal I_3$, a closed formula for the SW dispersion can be obtained via hydrodynamic or linear spin wave theory~\cite{Halperin1969}. In this theory, the localized spins in Eq.~\eqref{eq:LLGmodified}, \mbox{$\mathbf{M}_{n} = ({\rm Re}\, \phi_{n}, {\rm Im}\, \phi_{n}, 1-m)^T$}, are expressed using complex field $\phi_{n}$ and uniform spin density $m\ll 1$. Then, using the SW ansatz $\phi_{n}(t) = \sum_\mathbf{q} U_\mathbf{q} e^{i(\mathbf{q}\cdot\mathbf{r}_n-\omega_\mathbf{q} t)}$, we obtain the dispersion relation for the SWs
\begin{equation}\label{eq:dispersion}
    \omega_\mathbf{q} = (Jq^2 + K - B)\big[1 + i(\alpha_G + \tilde{\lambda}_\mathbf{q})\big],
\end{equation}
where $\mathbf{q}$ is the wavevector and $\omega$ is their frequency. The damping of the SW in the linear regime is then given by the imaginary part of the dispersion in Eq.~\eqref{eq:dispersion}, \mbox{$\Gamma_\mathbf{q}^{\rm linear}\equiv {\rm Im}\,\omega_\mathbf{q}$}. It is comprised by contributions from the local scalar Gilbert damping $\alpha_G$ and the Fourier transform of the nonlocal damping tensor, $\tilde{\lambda}_\mathbf{q} = \int\!d\mathbf{r}_n \,\lambda_{\mathbf{r}_n} e^{i\mathbf{q\cdot r}_n}$.

\section{Results for FI/NM bilayer}

We warm up by extracting $\Gamma_\mathbf{q}$ for the simplest of the three cases in Fig.~\ref{fig:setup}, a one-dimensional (1D) FI chain under a 1D NM overlayer with spin-degenerate quadratic electronic energy-momentum dispersion, \mbox{$\epsilon_{\mathbf{k\sigma}} = t_0k_x^2$}, where $t_0=\hbar^2/2m$. The GFs and spectral functions in Eq.~\eqref{eq:nonlocalDamping}, can be calculated in the momentum representation, yielding $\lambda_{R}^{\rm 1D} = \frac{2J_{sd}^2}{\pi v_F^2}\cos^2(k_F R)\mathcal{I}_3$, where $v_F$ is the Fermi velocity, $R\equiv|\mathbf{R}|$, and $k_F$ is the Fermi wavevector. Moreover, its Fourier transform, \mbox{$\tilde\lambda_q = \frac{2J_{sd}^2}{v_F^2} [\delta(q) + \delta(q-2k_F)/2]$}, dictates additional damping to SWs of wavevector $q=0, \pm 2k_F$. Although the Dirac delta function in this expression is unbounded, this unphysical feature is an artifact of the small amplitude, $m\ll1$, approximation within the hydrodynamic approach~\cite{Halperin1969}. The features of such wavevector-dependent damping in 1D can be \textit{corroborated} via TDNEGF+LLG (TDNEGF stands for time-dependent nonequilibrium GF~\cite{Gaury2014})  numerically exact simulations~\cite{Bajpai2019,Petrovic2018,Petrovic2021,Suresh2020} of a finite-size nanowire, similar to the setup depicted in Fig.~\ref{fig:setup}(a) but sandwiched between two NM semi-infinite leads. For example, by exciting SWs via injection of spin-polarized current into the metallic overlayer of such a system, as pursued experimentally in spintronics and magnonics~\cite{Madami2011, Demidov2020}, we find in Fig.~\ref{fig:1Ddamping}(a) that wavevector $q$ of thereby excited coherent SW increases with increasing anisotropy $K$. However, the maximum wavevector $q_{\rm max}$ is limited by $k_F$ [Fig.~\ref{fig:1Ddamping}(b)]. This means that SWs with $q \gtrsim k_F$ are subjected to additional damping, inhibiting their generation. Although our analytical results predict extra damping at $q=2k_F$, finite size effects and the inclusion of semi-infinite leads in TDNEGF+LLG simulations lower this cutoff to $k_F$.

Since SW experiments are conducted on higher-dimensional systems, we also investigate damping on SWs in a two-dimensional (2D) FI/NM bilayer. The electronic energy-momentum dispersion is then \mbox{$\epsilon_{\mathbf{k\sigma}} = t_0(k_x^2 + k_y^2)$}, and the nonlocal damping and its Fourier transform are given by 
\begin{eqnarray}
\lambda_{R}^{\rm NM} &=& \frac{k_F^2J_{sd}^2}{2\pi v_F^2}J_0^2(k_F R)\mathcal{I}_3 \label{eq:damping2dSpace}, \\
    \tilde\lambda_q^{\rm NM} &=& \frac{k_F J_{sd}^2  \Theta(2k_F - q)}{2\pi v_F^2 q\sqrt{1-(q/2k_F)^2}},\label{eq:damping2dWave}
\end{eqnarray}
where $J_n(x)$ is the $n$-th Bessel function of the first kind, and $\Theta(x)$ is the Heaviside step function. The nonlocal damping in Eqs.~\eqref{eq:damping2dSpace} and~\eqref{eq:damping2dWave} is plotted in Fig.~\ref{fig:2Ddamping}(a), showing realistic decay with increasing $R$, in contrast to unphysical infinite range found in 1D case. Additionally, SW damping in Eq.~\eqref{eq:damping2dWave} is operative for wavectors $0\leq q\leq 2k_F$, again diverging for $q=0,2k_F$ due to artifacts of hydrodynamic theory~\cite{Halperin1969}. Therefore, unphysical divergence can be removed by going back to discrete lattice and solving numerically a system of coupled LLG Eq.~\eqref{eq:LLGmodified} where $\lambda_\mathbf{R}$ in 2D is used~\cite{ReyesOsorio2023a}. From the exponential decay of the initial SW amplitude in such numerical solutions, we extract the wavevector-dependent SW damping $\Gamma_\mathbf{q}$, plotted as as solid curves in Fig.~\ref{fig:2Ddamping}(e). Note that SW damping obtained in this fashion includes nonlinear effects such as magnon-magnon interaction. In this numerical treatment we use $n=$1--100 LMMs; $k_F = 0.5a^{-1}$ where $a$ is the lattice spacing; $k_F^2J_{sd}^2/2\pi v_F^2=\eta=0.1$; $K=0$; $B_{\rm ext}=0.1\, J$; and $\alpha_G=0.1$. 

\section{Results for FI/HM bilayer}

\begin{figure}
    \centering
    \includegraphics[width=\columnwidth]{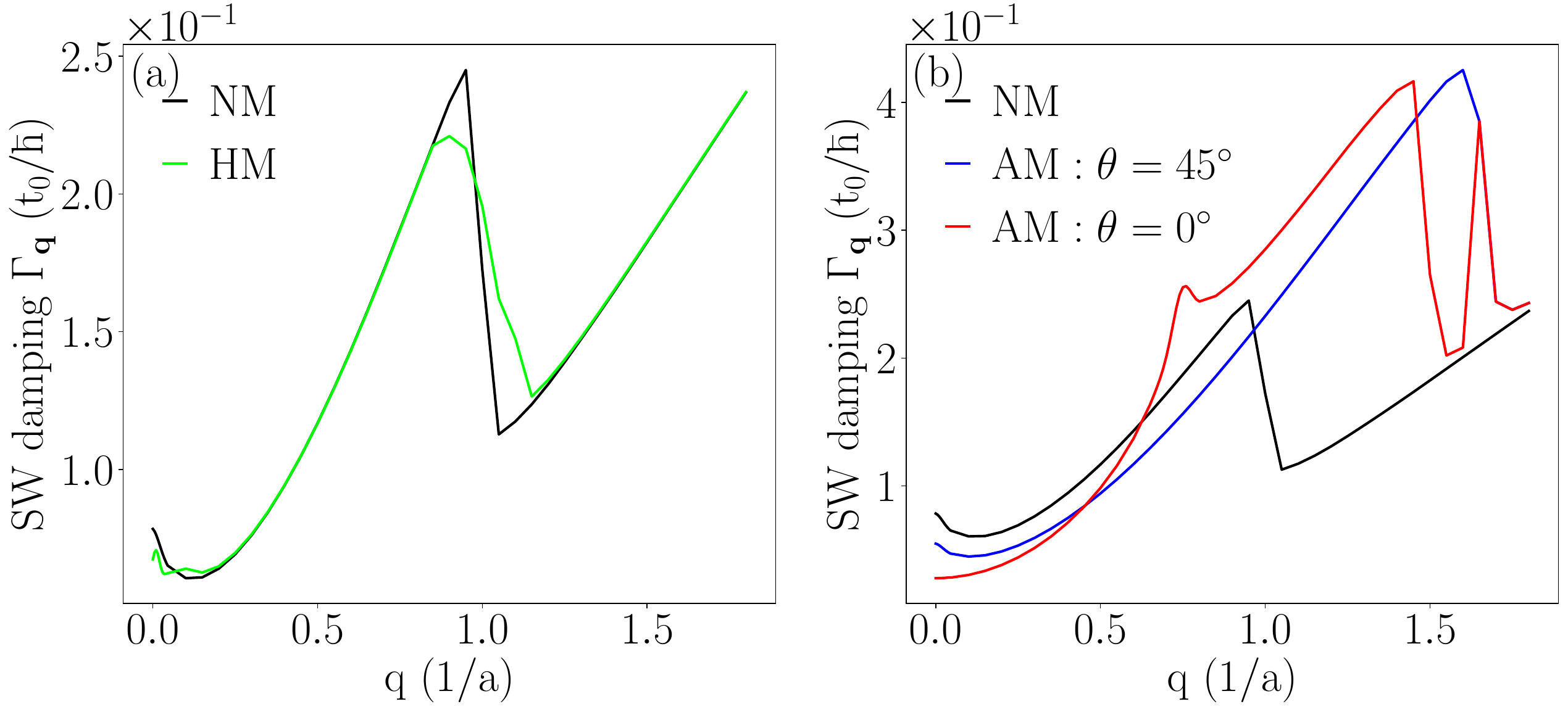}
    \caption{(a) Wavevector-dependent damping $\Gamma_\mathbf{q}$ of SWs under NM or HM overlayer with the Rashba SOC of strength $t_{\rm SOC}=0.1t_0$. (b) $\Gamma_\mathbf{q}$ of SWs under AM overlayer with $t_{\rm AM}=0.8t_0$ and for different relative orientations of FI and AM layers measured by angle $\theta$ [Fig.~\ref{fig:setup}]. All calculations employ $\eta=0.1$ and Fermi energy $\varepsilon_F=0.25t_0$.}
    \label{fig:SWdamping}
\end{figure}

Heavy metals (such as often employed Pt, W, Ta) exhibit strong SOC effects due to their large atomic number. We mimic their presence at the FI/HM interface~\cite{Bihlmayer2022} by using 2D energy-momentum dispersion $\epsilon_{\mathbf{k}} = t_0 (k^2_x + k^2_y) + t_{\rm SOC}(\sigma^x k_y - \sigma^y k_x)$, which includes spin-splitting due to the Rashba SOC~\cite{Bihlmayer2022, Manchon2015}. Using this dispersion, Eq.~\eqref{eq:nonlocalDamping} yields
\begin{equation}\label{eq:HMdampingTensor}
    \lambda_\mathbf{R}^{\rm HM} = \begin{pmatrix}
        \lambda^{xx}_\mathbf{R} & \lambda^{xy}_\mathbf{R} & 0 \\
        \lambda^{xy}_\mathbf{R} & \lambda^{yy}_\mathbf{R} & 0 \\
        0 & 0 & \lambda^{zz}_\mathbf{R}
    \end{pmatrix},
\end{equation}
for the nonlocal damping tensor. Its components are, in general, different from each other
\begin{widetext}
\begin{subequations}\label{eq:HMdamping}
    \begin{align}
        \lambda^{xx}_\mathbf{R} &= \frac{J_{sd}^2}{4\pi}\bigg[\bigg(\frac{k_{F\uparrow}}{v_{F\uparrow}}J_0(k_{F\uparrow} R) + \frac{k_{F\downarrow}}{v_{F\downarrow}}J_0(k_{F\downarrow} R)\bigg)^2 + \cos(2\theta) \bigg(\frac{k_{F\uparrow}}{v_{F\uparrow}}J_1(k_{F\uparrow} R) - \frac{k_{F\downarrow}}{v_{F\downarrow}}J_1(k_{F\downarrow} R)\bigg)^2 \bigg], \label{eq:HMdampingxx}\\
        \lambda^{yy}_\mathbf{R} &= \frac{J_{sd}^2}{4\pi}\bigg[\bigg(\frac{k_{F\uparrow}}{v_{F\uparrow}}J_0(k_{F\uparrow} R) + \frac{k_{F\downarrow}}{v_{F\downarrow}}J_0(k_{F\downarrow} R)\bigg)^2 - \cos(2\theta) \bigg(\frac{k_{F\uparrow}}{v_{F\uparrow}}J_1(k_{F\uparrow} R) - \frac{k_{F\downarrow}}{v_{F\downarrow}}J_1(k_{F\downarrow} R)\bigg)^2 \bigg], \label{eq:HMdampingyy}\\
        \lambda^{zz}_\mathbf{R} &= \frac{J_{sd}^2}{4\pi}\bigg[\bigg(\frac{k_{F\uparrow}}{v_{F\uparrow}}J_0(k_{F\uparrow} R) + \frac{k_{F\downarrow}}{v_{F\downarrow}}J_0(k_{F\downarrow} R)\bigg)^2 - \bigg(\frac{k_{F\uparrow}}{v_{F\uparrow}}J_1(k_{F\uparrow} R) - \frac{k_{F\downarrow}}{v_{F\downarrow}}J_1(k_{F\downarrow} R)\bigg)^2 \bigg], \label{eq:HMdampingzz}\\
        \lambda^{xy}_\mathbf{R} &= -\frac{J_{sd}^2 \sin(2\theta)}{4\pi} \bigg(\frac{k_{F\uparrow}}{v_{F\uparrow}}J_1(k_{F\uparrow} R) - \frac{k_{F\downarrow}}{v_{F\downarrow}}J_1(k_{F\downarrow} R)\bigg)^2, \label{eq:HMdampingxy}
    \end{align}
\end{subequations}
\end{widetext}
where $k_{F\uparrow}$ and $k_{F\downarrow}$ are the spin-split Fermi wavevectors [Fig.~\ref{fig:setup}(b)], and $\theta$ is the relative orientation angle [Fig.~\ref{fig:setup}(b)] between the SW wavevector $\mathbf q$ and the $k_x$ direction. Thus, the nonlocal damping tensor in Eq.~\eqref{eq:HMdampingTensor}  generated by HM overlayer is anisotropic in spin due to its different diagonal elements, as well as nonzero off-diagonal elements. It is also anisotropic in space due to its dependence on the angle $\theta$. Its elements [Eqs.~\eqref{eq:HMdamping}] are plotted in Figs.~\ref{fig:2Ddamping}(b),~\ref{fig:2Ddamping}(c),~\ref{fig:2Ddamping}(f), and~\ref{fig:2Ddamping}(g) using $t_{\rm SOC}=0.3 t_0$, and reduce to the FI/NM case if $t_{\rm SOC}=0$. They may become negative, signifying the possibility of antidamping torque~\cite{Demidov2020} exerted by conduction electrons. However, the dominant effect of nearby LMMs and the presence of local scalar $\alpha_G$ ensures that LMM dynamics is damped overall. Although there is no closed expression for the SW dispersion in the presence of anisotropic $\lambda_\mathbf{R}^{\rm HM}$, we can still extract SW damping $\Gamma_\mathbf{q}$ induced by an HM overlayer from the exponential decay of the SW amplitude in numerical integration of extended LLG Eq.~\eqref{eq:LLGmodified} using SW initial conditions with varying $\mathbf{q}$. For an HM overlayer with realistic~\cite{Bihlmayer2022, Manchon2015} $t_{\rm SOC} = 0.1 t_0$ the results in Fig.~\eqref{fig:SWdamping}(a) are very similar to those obtained for NM overlayer with the same Fermi energy. Also, the spatial anisotropy of $\lambda_\mathbf{R}^{\rm HM}$ did not translate into $\theta$-dependence of the SW damping.

\section{Results for FI/AM bilayer} 

Altermagnets~\cite{Smejkal2022b, Smejkal2022a} are a novel class of antiferromagnets with spin-split electronic energy-momentum dispersion despite zero net magnetization or lack of SOC. They are currently intensely explored as a new resource for spintronics~\cite{Sun2023b, Bai2023, Zhang2024} and magnonics~\cite{Smejkal2022c, Jin2023}. A simple model for an AM overlayer employs energy-momentum dispersion \mbox{$\epsilon_{\mathbf{k}\sigma} = t_0(k_x^2 + k_y^2) - t_{\rm AM}\sigma(k_x^2 - k_y^2)$}~\cite{Smejkal2022a, Smejkal2022b}, where $t_{\rm AM}$ is the parameter characterizing anisotropy in the AM. The corresponding \mbox{$\lambda_\mathbf{R}^{\rm AM} = {\rm diag}(\lambda_\mathbf{R}^\perp, \lambda_\mathbf{R}^\perp, \lambda_\mathbf{R}^\parallel)$} tensor has three components, which we derive from Eq.~\eqref{eq:nonlocalDamping} as
\begin{subequations}\label{eq:AMdamp}
    \begin{align}
        \lambda_\mathbf{R}^\perp &= \frac{J_{sd}^2}{4\pi A_+ A_-}\bigg[J_0^2\bigg(\sqrt{\frac{\epsilon_F}{t_0}}R_+\bigg) + J_0^2\bigg(\sqrt{\frac{\epsilon_F}{t_0}}R_-\bigg)\bigg], \label{eq:AMdampPerp}\\
        \lambda_\mathbf{R}^\parallel &= \frac{J_{sd}^2}{2\pi A_+ A_-} J_0\bigg(\sqrt{\frac{\epsilon_F}{t_0}}R_+\bigg)J_0\bigg(\sqrt{\frac{\epsilon_F}{t_0}}R_-\bigg), \label{eq:AMdampPar}
    \end{align}
\end{subequations}
where $A_\pm = t_0 \pm t_{\rm AM}$ and $R_\pm^2 = X^2/A_\pm + Y^2/A_\mp$ is the anisotropically rescaled norm of $\mathbf{R}$. They are plotted in Figs.~\ref{fig:2Ddamping}(d) and~\ref{fig:2Ddamping}(h), demonstrating that $\lambda_\mathbf{R}^{\rm AM}$ is highly anisotropic in space and spin due to the importance of angle $\theta$~\cite{Papaj2023, Sun2023a, Sun2023b}. Its components can also take  negative values, akin to the case of $\lambda^{\rm HM}_\mathbf{R}$. It is interesting to note that along the direction of $\theta = 45^\circ$ [gray dashed line in Figs.~\ref{fig:2Ddamping}(d) and~\ref{fig:2Ddamping}(h)], $\lambda^\perp_\mathbf{R} = \lambda_\mathbf{R}^\parallel$ so that nonlocal damping tensor is isotropic in spin. The SW damping $\Gamma_\mathbf{q}$ induced by an AM overlayer is extracted from numerical integration of extended LLG Eq.~\eqref{eq:LLGmodified} and plotted in Fig.~\eqref{fig:SWdamping}(b). Using a relatively large, but realistic~\cite{Smejkal2022b}, AM parameter $t_{\rm AM}=0.8t_0$, the SW damping for experimentally relevant small wavevectors is reduced when compared to the one due to NM overlayer by up to 65\% for $\theta=0^\circ$ [Fig.~\ref{fig:SWdamping}(b)]. Additional nontrivial features are observed at higher $|\mathbf{q}|$, such as being operative for a greater range of wavevectors and with maxima around $|\mathbf{q}|=2\sqrt{\epsilon_F/t_0}$ and $|\mathbf{q}|=3\sqrt{\epsilon_F/t_0}$. Remarkably, these peaks vanish for wavevectors along the isotropic direction $\theta=45^\circ$ [Fig.~\ref{fig:SWdamping}(b)].

\section{Conclusions}

In conclusion, using SKFT-derived nonlocal damping tensor~\cite{ReyesOsorio2023a}, we demonstrated a rigorous path to obtain wavevector damping of SWs in magnetic insulator due to interaction with conduction electrons of metallic overlayer, as a setup often encountered in magnonics~\cite{Chumak2015, Serga2010, Chumak2019, Chumak2022, Mahmoud2020, Csaba2017} where such SW damping was directly measured in very recent experiments~\cite{Bertelli2021, Mae2022, Krysztofik2022, Serha2022}. Our analytical expressions [Eqs.~\eqref{eq:damping2dSpace},~\eqref{eq:HMdampingTensor}, and~\eqref{eq:AMdamp}] for nonlocal damping tensor---using simple models of NM, HM, and AM overlayers as an input---can be directly plugged into atomistic spin dynamics simulations~\cite{Evans2014}. For more complicated band structures of metallic overlayers, one can compute $\lambda_\mathbf{R}$ numerically via Eq.~\eqref{eq:nonlocalDamping}, including combination with first-principles calculations~\cite{Lu2023}. 

The interfacial $J_{sd}$ coupling has been estimated to be 50 meV in YIG/Pt~\cite{Troncoso2019}, and as high as 400 meV in YIG/Au~\cite{Rohling2018}, accounting for the 100-fold increase~\cite{Bertelli2021, Mae2022, Krysztofik2022, Serha2022} in SW damping underneath a metallic overlayer (see the SM~\cite{Note1} for more details). This increase is incompatible with prior classical theories of nonlocal damping~\cite{Kapelrud2013, Skarsvaag2014}, emphasizing the necessity of the quantum mechanical foundations of our SKFT-based approach. Since we find that $\lambda_\mathbf{R}^{\rm HM,AM}$ is highly anisotropic in spin and space, the corresponding SW damping $\Gamma_\mathbf{q}$ thus understood microscopically from SKFT allows us to propose how to manipulate it [Fig.~\ref{fig:SWdamping}]. For example, by using HM or AM overlayer and by changing their relative orientation with respect to FI layer, $\Gamma_\mathbf{q}$ can be reduced by up to 65\% at small wavevectors $\mathbf{q}$ [Fig.~\ref{fig:SWdamping}], which can be of great interest to magnonics experiments and applications.

\appendix
\section{Estimating Magnitude of SW damping for realistic YIG/Au bilayer from SKFT vs. from Ref.~\cite{Kapelrud2013}}\label{app:a}

The strength of the SKFT-derived nonlocal damping that gives rise to the wavevector-dependent damping of SWs is given by the dimensionless parameter
\begin{equation}\label{eq:eta}
\eta=\frac{k_F^2 J_{sd}^2 a_{\rm Au}^4}{2\pi \hbar^2 v_F^2},
\end{equation}
where $J_{sd}$ is the interfacial $sd$ exchange coupling; $k_F$ and $v_F$ are the norm of the Fermi wavevector and Fermi velocity, respectively; and $a_{\rm Au}$ is the distance between nearest-neighbors (NN) in Au. Reference~\cite{Rohling2018}, based on a many-body model, estimate the $J_{sd}$ coupling for a YIG/Au bilayer to be given by
\begin{equation}\label{eq:jsd}
    J_{sd} = 0.33 \cdot 2\pi t_0 a_{\rm Au} \frac{\sqrt{g_{\uparrow\downarrow}}}{s},
\end{equation}
where $t_0$ is the nearest-neighbor (NN) hopping parameter of Au; $g_{\uparrow\downarrow}$ is the spin-mixing conductance at the interface; and $s$ is the spin quantum number of YIG. The latter is estimated to be $s=M_s a_{\rm YIG} a_{\rm Au}^2/\hbar\gamma_e$, where \mbox{$M_s = 1.6\times 10^5$ A/m} is the saturation magnetization of YIG; \mbox{$a_{\rm YIG}=12$ \r A} is the lattice constant of YIG; and \mbox{$\gamma_e = 1.76\times10^{11}\ {\rm rad.s^{-1}T^{-1}}$} is the gyromagnetic ratio of the electron. The NN distance in Au is $a_{\rm Au} = \sqrt{0.63/g_{\rm sh}}$, where $g_{\rm sh}=12 \ {\rm nm^{-2}}$ is the Sharvin conductance of Au. Therefore, the spin quantum number of YIG is approximately $s=0.54$, or about spin $1/2$.

Next, the Au hopping parameter is estimated to be $t_0=\varepsilon_F/12$, where $\varepsilon_F=5.5$ eV is the Fermi energy of Au. This allows us to determine $t_0=0.46$ eV, as well as the norm of the Fermi wavevector $k_F=10.69 \ {\rm nm^{-1}}$ and Fermi velocity $1.6\times 10^{6}$ m/s. The spin-mixing conductance at the interface, $g_{\uparrow\downarrow}$, is reported~\cite{Rohling2018} to be between 1.2--6 nm$^{-1}$: we use the lower bound of $g_{\uparrow\downarrow}=1.2$ nm$^{-1}$. Plugging into Eq.~\eqref{eq:jsd}, we get a lower bound for $J_{sd} = 0.44$ eV. Then, in turn, plugging into Eq.~\eqref{eq:eta} yields $\eta=8.6\times 10^{-3}$. Let us recall that the local scalar Gilbert damping of YIG is $\sim 10^{-4}$, meaning that the effective damping under an Au overlayer is predicted to be \textit{at least }$\sim 80$ times stronger than in pure YIG. This is fully compatible with recent NV magnetometry measurements that observe a \textit{100-fold} increase. Contrariwise, some theories based on classical electrodynamics~\cite{Kapelrud2013, Skarsvaag2014} predict a modest $\sim 2$\textit{-fold} increase. Crucially, these References employ the same spin-mixing conductance of $g_{\uparrow\downarrow}=1.2$ nm$^{-1}$ as the key quantity to obtain the result.

\section{Estimating Magnitude of SW damping due to backaction of eddy currents in Au generated by dynamical magnetic moments of YIG}

\begin{figure}
    \centering
    \includegraphics[width=\linewidth]{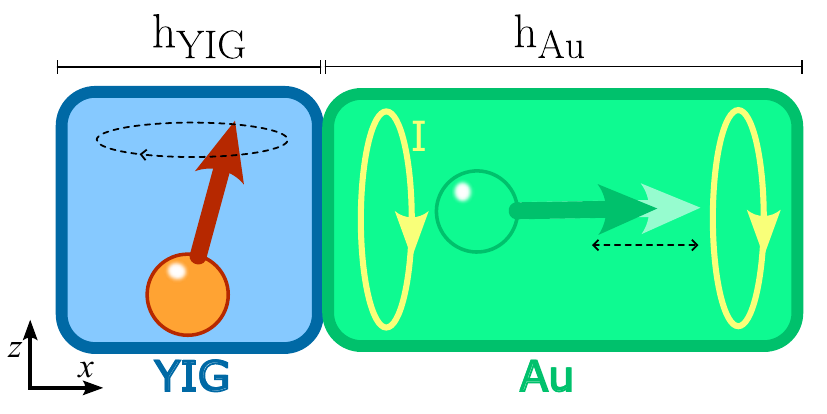}
    \caption{Schematic diagram of a classical precessing localized magnetic moment in YIG layer that induces eddy currents $I$ within an Au overlayer, as used in the theory of Ref.~\cite{Bertelli2021}.}
    \label{fig:classical}
\end{figure}

The NV magnetometry experiments showing the dramatic increase in SW damping~\cite{Bertelli2021} employ a classical electrodynamics framework in which precessing magnetic moments in YIG induce eddy currents in the Au overlayer. The eddy currents, in turn, generate a magnetic moment whose magnetic field exerts a damping-like torque on the localized magnetic moments of the YIG. However, Ref.~\cite{Bertelli2021} leave the damping as a free parameter. Here, we compute the damping parameter of a single classical precessing moment based on realistic material parameters. The geometry we consider is depicted in Fig.~\ref{fig:classical}: a cube of YIG of thickness $h_{\rm YIG}=235$ nm under a layer of Au of thickness $h_{\rm Au}=200$ nm. The magnetization of YIG cube is averaged to determine a single magnetic dipole moment $m_{\rm YIG}=M_s h_{\rm YIG}^3$ that exhibits precessing motion according to $\mathbf{m}_{\rm YIG} = m_{\rm YIG}(\sin\theta\sin\omega t, \sin\theta\cos\omega t, \cos\theta)$. The precession frequency is $\omega/2\pi= 2$ GHz and $\theta$ is the angle of the magnetic moment with the $z$ axis. 

The magnetic field generated by a localized magnetic dipole~\cite{Griffiths2011} is given by
\begin{equation}\label{eq:field}
    \mathbf{B}(\mathbf{r}) = \frac{\mu_0}{4\pi}\left[ \frac{\mathbf{e}_r(\mathbf{e}_r \cdot \partial_t^2\mathbf{m}) - \partial_t^2\mathbf{m}}{rc^2} + \frac{3\mathbf{e}_r(\mathbf{e}_r \cdot \mathbf{M}) - \mathbf{M}}{r^3} \right],
\end{equation}
where $c$ is the speed of light; $\mathbf{e}_r$ is the radial unit vector at position $\mathbf{r}$;  $\mathbf{M}=\mathbf{m}+r\partial_t\mathbf{m}/c$, and retardation effects can be ignored since $\omega h_{\rm Au}/c\sim 10^{-6}$. Assuming the Au layer to be thin enough, only the $x$-component of the magnetic field radiated by $\mathbf{m}_{\rm YIG}$, $B_{\rm YIG}^x$, contributes to the flux through the Au. This component is obtained from Eq.~\eqref{eq:field} as
\begin{equation}\label{eq:fieldyig}
    B^x_{\rm YIG} (x, y=0,z=0) = \frac{\mu_0 m_{\rm YIG}}{2\pi x^3}\sin\theta\cos\omega t,
\end{equation}
where terms of order $\omega x/c$ are neglected. Thus, the electromotive force $\epsilon$ induced in the Au is given by
\begin{equation}
\epsilon =  h_{\rm YIG}^2\int_{\rm Au} \! dx \, \partial_t B^x_{\rm YIG} (x, y=0,z=0),
\end{equation}
where the flux is integrated over the thickness of the Au layer. Thus, the circulating eddy currents are $I=\epsilon/\rho_{\rm Au} = I_0\sin\omega t$, where $\rho_{\rm Au}=11$ n$\Omega$m is the resistivity of the Au thin film at room temperature~\cite{Sambles1982}. Such currents generate an induced magnetic dipole $\mathbf{m}_{\rm Au} = I_0 h_{\rm YIG}^2 \sin\omega t \mathbf{e}_x$, whose generated magnetic field at the location of the YIG magnetic moment is obtained from Eq.~\eqref{eq:field} as
\begin{equation}
    \mathbf{B}_{\rm Au} = \frac{4\mu_0 I_0 h_{\rm YIG}^2 \sin\omega t }{\pi (h_{\rm Au} + h_{\rm YIG})^3} \mathbf{e}_x.
\end{equation}

The torque exerted by this field can be included in the LLG equations for a precessing magnetic moment as an additional term, yielding
\begin{align}\label{eq:llg}
    \partial_t \mathbf{m}_{\rm YIG} &= -\gamma_e \mathbf{m}_{\rm YIG} \times \mathbf{B}_{\rm ext} + \alpha_G \mathbf{m}_{\rm YIG} \times \partial_t \mathbf{m}_{\rm YIG} \\
    &+ \alpha_1 \mathbf{m}_{\rm YIG} \times (\partial_t \mathbf{m}_{\rm YIG} \cdot \mathbf{e}_x)\mathbf{e}_x,\nonumber 
\end{align}
where $\alpha_G$ is the conventional Gilbert damping scalar, and $\alpha_1 = \gamma_e B_{\rm Au}/\omega$ is the additional damping due to the eddy currents. This parameter is, approximately, $\alpha_1 \approx 0.012$, compatible with the experimental observations~\cite{Bertelli2021} and \textit{100-fold} larger than intrinsic $\alpha_G=10^{-4}$. However, Eq.~\eqref{eq:llg} is the usual LLG equation with traditional uniform Gilbert damping whose value is anisotropically enhanced [second term on the right-hand side of Eq.~\eqref{eq:llg}] by the eddy-current backaction. It has been understood~\cite{Verba2018, Hankiewicz2008} that simple renormalization of Gilbert damping constant, which is the same for all SW modes at different wavevectors, is insufficient to produce wavevector-dependent damping. Instead, wavevector-dependent SW damping requires nonlocal damping term in the LLG equation, which is coordinate- and magnetization-texture-dependent. Deriving such nonlocal damping term cannot be achieved using simple physical arguments leading to Eq.~\eqref{eq:llg}, instead one needs more microscopic approaches~\cite{Hankiewicz2008,Tserkovnyak2009,Zhang2009,Yuan2016,Verba2018}, one of which (and perhaps most rigorous due to being benchmarked by numerically exact calculations~\cite{ReyesOsorio2023a, Sayad2015}) is provided by our SKFT-based approach.

\bibliography{master-biblio}

\begin{thebibliography}{75}%
\makeatletter
\providecommand \@ifxundefined [1]{%
 \@ifx{#1\undefined}
}%
\providecommand \@ifnum [1]{%
 \ifnum #1\expandafter \@firstoftwo
 \else \expandafter \@secondoftwo
 \fi
}%
\providecommand \@ifx [1]{%
 \ifx #1\expandafter \@firstoftwo
 \else \expandafter \@secondoftwo
 \fi
}%
\providecommand \natexlab [1]{#1}%
\providecommand \enquote  [1]{``#1''}%
\providecommand \bibnamefont  [1]{#1}%
\providecommand \bibfnamefont [1]{#1}%
\providecommand \citenamefont [1]{#1}%
\providecommand \href@noop [0]{\@secondoftwo}%
\providecommand \href [0]{\begingroup \@sanitize@url \@href}%
\providecommand \@href[1]{\@@startlink{#1}\@@href}%
\providecommand \@@href[1]{\endgroup#1\@@endlink}%
\providecommand \@sanitize@url [0]{\catcode `\\12\catcode `\$12\catcode `\&12\catcode `\#12\catcode `\^12\catcode `\_12\catcode `\%12\relax}%
\providecommand \@@startlink[1]{}%
\providecommand \@@endlink[0]{}%
\providecommand \url  [0]{\begingroup\@sanitize@url \@url }%
\providecommand \@url [1]{\endgroup\@href {#1}{\urlprefix }}%
\providecommand \urlprefix  [0]{URL }%
\providecommand \Eprint [0]{\href }%
\providecommand \doibase [0]{https://doi.org/}%
\providecommand \selectlanguage [0]{\@gobble}%
\providecommand \bibinfo  [0]{\@secondoftwo}%
\providecommand \bibfield  [0]{\@secondoftwo}%
\providecommand \translation [1]{[#1]}%
\providecommand \BibitemOpen [0]{}%
\providecommand \bibitemStop [0]{}%
\providecommand \bibitemNoStop [0]{.\EOS\space}%
\providecommand \EOS [0]{\spacefactor3000\relax}%
\providecommand \BibitemShut  [1]{\csname bibitem#1\endcsname}%
\let\auto@bib@innerbib\@empty
\bibitem [{\citenamefont {Li}\ and\ \citenamefont {Bailey}(2016)}]{Li2016}%
  \BibitemOpen
  \bibfield  {author} {\bibinfo {author} {\bibfnamefont {Y.}~\bibnamefont {Li}}\ and\ \bibinfo {author} {\bibfnamefont {W.}~\bibnamefont {Bailey}},\ }\bibfield  {title} {\bibinfo {title} {Wave-number-dependent {Gilbert} damping in metallic ferromagnets.},\ }\href {https://doi.org/10.1103/PhysRevLett.116.117602} {\bibfield  {journal} {\bibinfo  {journal} {Phys. Rev. Lett.}\ }\textbf {\bibinfo {volume} {116}},\ \bibinfo {pages} {117602} (\bibinfo {year} {2016})}\BibitemShut {NoStop}%
\bibitem [{\citenamefont {Chen}\ \emph {et~al.}(2022)\citenamefont {Chen}, \citenamefont {Mao}, \citenamefont {Chung}, \citenamefont {Stone}, \citenamefont {Kolesnikov}, \citenamefont {Wang}, \citenamefont {Murai}, \citenamefont {Gao}, \citenamefont {Delaire},\ and\ \citenamefont {Dai}}]{Chen2022}%
  \BibitemOpen
  \bibfield  {author} {\bibinfo {author} {\bibfnamefont {L.}~\bibnamefont {Chen}}, \bibinfo {author} {\bibfnamefont {C.}~\bibnamefont {Mao}}, \bibinfo {author} {\bibfnamefont {J.-H.}\ \bibnamefont {Chung}}, \bibinfo {author} {\bibfnamefont {M.~B.}\ \bibnamefont {Stone}}, \bibinfo {author} {\bibfnamefont {A.~I.}\ \bibnamefont {Kolesnikov}}, \bibinfo {author} {\bibfnamefont {X.}~\bibnamefont {Wang}}, \bibinfo {author} {\bibfnamefont {N.}~\bibnamefont {Murai}}, \bibinfo {author} {\bibfnamefont {B.}~\bibnamefont {Gao}}, \bibinfo {author} {\bibfnamefont {O.}~\bibnamefont {Delaire}},\ and\ \bibinfo {author} {\bibfnamefont {P.}~\bibnamefont {Dai}},\ }\bibfield  {title} {\bibinfo {title} {Anisotropic magnon damping by zero-temperature quantum fluctuations in ferromagnetic {CrGeTe$_3$}},\ }\href {https://doi.org/10.1038/s41467-022-31612-w} {\bibfield  {journal} {\bibinfo  {journal} {Nat. Commun.}\ }\textbf {\bibinfo {volume} {13}},\ \bibinfo {pages} {4037} (\bibinfo {year} {2022})}\BibitemShut {NoStop}%
\bibitem [{\citenamefont {Dai}\ \emph {et~al.}(2000)\citenamefont {Dai}, \citenamefont {Hwang}, \citenamefont {Zhang}, \citenamefont {Fernandez-Baca}, \citenamefont {Cheong}, \citenamefont {Kloc}, \citenamefont {Tomioka},\ and\ \citenamefont {Tokura}}]{Dai2000}%
  \BibitemOpen
  \bibfield  {author} {\bibinfo {author} {\bibfnamefont {P.}~\bibnamefont {Dai}}, \bibinfo {author} {\bibfnamefont {H.~Y.}\ \bibnamefont {Hwang}}, \bibinfo {author} {\bibfnamefont {J.}~\bibnamefont {Zhang}}, \bibinfo {author} {\bibfnamefont {J.~A.}\ \bibnamefont {Fernandez-Baca}}, \bibinfo {author} {\bibfnamefont {S.-W.}\ \bibnamefont {Cheong}}, \bibinfo {author} {\bibfnamefont {C.}~\bibnamefont {Kloc}}, \bibinfo {author} {\bibfnamefont {Y.}~\bibnamefont {Tomioka}},\ and\ \bibinfo {author} {\bibfnamefont {Y.}~\bibnamefont {Tokura}},\ }\bibfield  {title} {\bibinfo {title} {Magnon damping by magnon-phonon coupling in manganese perovskites},\ }\href {https://doi.org/10.1103/PhysRevB.61.9553} {\bibfield  {journal} {\bibinfo  {journal} {Phys. Rev. B}\ }\textbf {\bibinfo {volume} {61}},\ \bibinfo {pages} {9553} (\bibinfo {year} {2000})}\BibitemShut {NoStop}%
\bibitem [{\citenamefont {Bayrakci}\ \emph {et~al.}(2013)\citenamefont {Bayrakci}, \citenamefont {Tennant}, \citenamefont {Leininger}, \citenamefont {Keller}, \citenamefont {Gibson}, \citenamefont {Wilson}, \citenamefont {Birgeneau},\ and\ \citenamefont {Keimer}}]{Bayrakci2013}%
  \BibitemOpen
  \bibfield  {author} {\bibinfo {author} {\bibfnamefont {S.~P.}\ \bibnamefont {Bayrakci}}, \bibinfo {author} {\bibfnamefont {D.~A.}\ \bibnamefont {Tennant}}, \bibinfo {author} {\bibfnamefont {P.}~\bibnamefont {Leininger}}, \bibinfo {author} {\bibfnamefont {T.}~\bibnamefont {Keller}}, \bibinfo {author} {\bibfnamefont {M.~C.~R.}\ \bibnamefont {Gibson}}, \bibinfo {author} {\bibfnamefont {S.~D.}\ \bibnamefont {Wilson}}, \bibinfo {author} {\bibfnamefont {R.~J.}\ \bibnamefont {Birgeneau}},\ and\ \bibinfo {author} {\bibfnamefont {B.}~\bibnamefont {Keimer}},\ }\bibfield  {title} {\bibinfo {title} {Lifetimes of antiferromagnetic magnons in two and three dimensions: Experiment, theory, and numerics},\ }\href {https://doi.org/10.1103/PhysRevLett.111.017204} {\bibfield  {journal} {\bibinfo  {journal} {Phys. Rev. Lett.}\ }\textbf {\bibinfo {volume} {111}},\ \bibinfo {pages} {017204} (\bibinfo {year} {2013})}\BibitemShut {NoStop}%
\bibitem [{\citenamefont {Zhitomirsky}\ and\ \citenamefont {Chernyshev}(2013)}]{Zhitomirsky2013}%
  \BibitemOpen
  \bibfield  {author} {\bibinfo {author} {\bibfnamefont {M.~E.}\ \bibnamefont {Zhitomirsky}}\ and\ \bibinfo {author} {\bibfnamefont {A.~L.}\ \bibnamefont {Chernyshev}},\ }\bibfield  {title} {\bibinfo {title} {{\em Colloquium}: Spontaneous magnon decays},\ }\href {https://doi.org/10.1103/revmodphys.85.219} {\bibfield  {journal} {\bibinfo  {journal} {Rev. Mod. Phys.}\ }\textbf {\bibinfo {volume} {85}},\ \bibinfo {pages} {219} (\bibinfo {year} {2013})}\BibitemShut {NoStop}%
\bibitem [{\citenamefont {Bajpai}\ \emph {et~al.}(2021)\citenamefont {Bajpai}, \citenamefont {Suresh},\ and\ \citenamefont {Nikoli{\'{c}}}}]{Bajpai2021}%
  \BibitemOpen
  \bibfield  {author} {\bibinfo {author} {\bibfnamefont {U.}~\bibnamefont {Bajpai}}, \bibinfo {author} {\bibfnamefont {A.}~\bibnamefont {Suresh}},\ and\ \bibinfo {author} {\bibfnamefont {B.~K.}\ \bibnamefont {Nikoli{\'{c}}}},\ }\bibfield  {title} {\bibinfo {title} {Quantum many-body states and {Green{\textquotesingle}s} functions of nonequilibrium electron-magnon systems: Localized spin operators versus their mapping to {Holstein}-{Primakoff} bosons},\ }\href {https://doi.org/10.1103/physrevb.104.184425} {\bibfield  {journal} {\bibinfo  {journal} {Phys. Rev. B}\ }\textbf {\bibinfo {volume} {104}},\ \bibinfo {pages} {184425} (\bibinfo {year} {2021})}\BibitemShut {NoStop}%
\bibitem [{\citenamefont {Smit}\ \emph {et~al.}(2020)\citenamefont {Smit}, \citenamefont {Keupert}, \citenamefont {Tsyplyatyev}, \citenamefont {Maksimov}, \citenamefont {Chernyshev},\ and\ \citenamefont {Kopietz}}]{Smit2020}%
  \BibitemOpen
  \bibfield  {author} {\bibinfo {author} {\bibfnamefont {R.~L.}\ \bibnamefont {Smit}}, \bibinfo {author} {\bibfnamefont {S.}~\bibnamefont {Keupert}}, \bibinfo {author} {\bibfnamefont {O.}~\bibnamefont {Tsyplyatyev}}, \bibinfo {author} {\bibfnamefont {P.~A.}\ \bibnamefont {Maksimov}}, \bibinfo {author} {\bibfnamefont {A.~L.}\ \bibnamefont {Chernyshev}},\ and\ \bibinfo {author} {\bibfnamefont {P.}~\bibnamefont {Kopietz}},\ }\bibfield  {title} {\bibinfo {title} {Magnon damping in the zigzag phase of the {Kitaev}-{Heisenberg}-$\mathrm{\ensuremath{\Gamma}}$ model on a honeycomb lattice},\ }\href {https://doi.org/10.1103/PhysRevB.101.054424} {\bibfield  {journal} {\bibinfo  {journal} {Phys. Rev. B}\ }\textbf {\bibinfo {volume} {101}},\ \bibinfo {pages} {054424} (\bibinfo {year} {2020})}\BibitemShut {NoStop}%
\bibitem [{\citenamefont {Winter}\ \emph {et~al.}(2017)\citenamefont {Winter}, \citenamefont {Riedl}, \citenamefont {Maksimov}, \citenamefont {Chernyshev}, \citenamefont {Honecker},\ and\ \citenamefont {Valent{\'i}}}]{Winter2017}%
  \BibitemOpen
  \bibfield  {author} {\bibinfo {author} {\bibfnamefont {S.~M.}\ \bibnamefont {Winter}}, \bibinfo {author} {\bibfnamefont {K.}~\bibnamefont {Riedl}}, \bibinfo {author} {\bibfnamefont {P.~A.}\ \bibnamefont {Maksimov}}, \bibinfo {author} {\bibfnamefont {A.~L.}\ \bibnamefont {Chernyshev}}, \bibinfo {author} {\bibfnamefont {A.}~\bibnamefont {Honecker}},\ and\ \bibinfo {author} {\bibfnamefont {R.}~\bibnamefont {Valent{\'i}}},\ }\bibfield  {title} {\bibinfo {title} {Breakdown of magnons in a strongly spin-orbital coupled magnet},\ }\href {https://doi.org/10.1038/s41467-017-01177-0} {\bibfield  {journal} {\bibinfo  {journal} {Nat. Commun.}\ }\textbf {\bibinfo {volume} {8}},\ \bibinfo {pages} {1152} (\bibinfo {year} {2017})}\BibitemShut {NoStop}%
\bibitem [{\citenamefont {Gohlke}\ \emph {et~al.}(2023)\citenamefont {Gohlke}, \citenamefont {Corticelli}, \citenamefont {Moessner}, \citenamefont {McClarty},\ and\ \citenamefont {Mook}}]{Gohlke2023}%
  \BibitemOpen
  \bibfield  {author} {\bibinfo {author} {\bibfnamefont {M.}~\bibnamefont {Gohlke}}, \bibinfo {author} {\bibfnamefont {A.}~\bibnamefont {Corticelli}}, \bibinfo {author} {\bibfnamefont {R.}~\bibnamefont {Moessner}}, \bibinfo {author} {\bibfnamefont {P.~A.}\ \bibnamefont {McClarty}},\ and\ \bibinfo {author} {\bibfnamefont {A.}~\bibnamefont {Mook}},\ }\bibfield  {title} {\bibinfo {title} {Spurious symmetry enhancement in linear spin wave theory and interaction-induced topology in magnons},\ }\href {https://doi.org/10.1103/PhysRevLett.131.186702} {\bibfield  {journal} {\bibinfo  {journal} {Phys. Rev. Lett.}\ }\textbf {\bibinfo {volume} {131}},\ \bibinfo {pages} {186702} (\bibinfo {year} {2023})}\BibitemShut {NoStop}%
\bibitem [{\citenamefont {Hankiewicz}\ \emph {et~al.}(2008)\citenamefont {Hankiewicz}, \citenamefont {Vignale},\ and\ \citenamefont {Tserkovnyak}}]{Hankiewicz2008}%
  \BibitemOpen
  \bibfield  {author} {\bibinfo {author} {\bibfnamefont {E.}~\bibnamefont {Hankiewicz}}, \bibinfo {author} {\bibfnamefont {G.}~\bibnamefont {Vignale}},\ and\ \bibinfo {author} {\bibfnamefont {Y.}~\bibnamefont {Tserkovnyak}},\ }\bibfield  {title} {\bibinfo {title} {Inhomogeneous {Gilbert} damping from impurities and electron-electron interactions},\ }\href {https://doi.org/10.1103/PhysRevB.78.020404} {\bibfield  {journal} {\bibinfo  {journal} {Phys. Rev. B}\ }\textbf {\bibinfo {volume} {78}},\ \bibinfo {pages} {020404(R)} (\bibinfo {year} {2008})}\BibitemShut {NoStop}%
\bibitem [{\citenamefont {Tserkovnyak}\ \emph {et~al.}(2009)\citenamefont {Tserkovnyak}, \citenamefont {Hankiewicz},\ and\ \citenamefont {Vignale}}]{Tserkovnyak2009}%
  \BibitemOpen
  \bibfield  {author} {\bibinfo {author} {\bibfnamefont {Y.}~\bibnamefont {Tserkovnyak}}, \bibinfo {author} {\bibfnamefont {E.~M.}\ \bibnamefont {Hankiewicz}},\ and\ \bibinfo {author} {\bibfnamefont {G.}~\bibnamefont {Vignale}},\ }\bibfield  {title} {\bibinfo {title} {Transverse spin diffusion in ferromagnets},\ }\href {https://doi.org/10.1103/physrevb.79.094415} {\bibfield  {journal} {\bibinfo  {journal} {Phys. Rev. B}\ }\textbf {\bibinfo {volume} {79}},\ \bibinfo {pages} {094415} (\bibinfo {year} {2009})}\BibitemShut {NoStop}%
\bibitem [{\citenamefont {Nikoli{\'{c}}}(2021)}]{Nikolic2021}%
  \BibitemOpen
  \bibfield  {author} {\bibinfo {author} {\bibfnamefont {P.}~\bibnamefont {Nikoli{\'{c}}}},\ }\bibfield  {title} {\bibinfo {title} {Universal spin wave damping in magnetic {Weyl} semimetals},\ }\href {https://doi.org/10.1103/physrevb.104.024414} {\bibfield  {journal} {\bibinfo  {journal} {Phys. Rev. B}\ }\textbf {\bibinfo {volume} {104}},\ \bibinfo {pages} {024414} (\bibinfo {year} {2021})}\BibitemShut {NoStop}%
\bibitem [{\citenamefont {Isoda}(1990)}]{Isoda1990}%
  \BibitemOpen
  \bibfield  {author} {\bibinfo {author} {\bibfnamefont {M.}~\bibnamefont {Isoda}},\ }\bibfield  {title} {\bibinfo {title} {Spin-wave damping in itinerant electron ferromagnets},\ }\href {https://doi.org/10.1088/0953-8984/2/15/014} {\bibfield  {journal} {\bibinfo  {journal} {J. Phys.: Condens. Matter}\ }\textbf {\bibinfo {volume} {2}},\ \bibinfo {pages} {3579} (\bibinfo {year} {1990})}\BibitemShut {NoStop}%
\bibitem [{\citenamefont {Buczek}\ \emph {et~al.}(2009)\citenamefont {Buczek}, \citenamefont {Ernst}, \citenamefont {Bruno},\ and\ \citenamefont {Sandratskii}}]{Buczek2009}%
  \BibitemOpen
  \bibfield  {author} {\bibinfo {author} {\bibfnamefont {P.}~\bibnamefont {Buczek}}, \bibinfo {author} {\bibfnamefont {A.}~\bibnamefont {Ernst}}, \bibinfo {author} {\bibfnamefont {P.}~\bibnamefont {Bruno}},\ and\ \bibinfo {author} {\bibfnamefont {L.~M.}\ \bibnamefont {Sandratskii}},\ }\bibfield  {title} {\bibinfo {title} {Energies and lifetimes of magnons in complex ferromagnets: A first-principle study of {Heusler} alloys},\ }\href {https://doi.org/10.1103/PhysRevLett.102.247206} {\bibfield  {journal} {\bibinfo  {journal} {Phys. Rev. Lett.}\ }\textbf {\bibinfo {volume} {102}},\ \bibinfo {pages} {247206} (\bibinfo {year} {2009})}\BibitemShut {NoStop}%
\bibitem [{\citenamefont {Chumak}\ \emph {et~al.}(2015)\citenamefont {Chumak}, \citenamefont {Vasyuchka}, \citenamefont {Serga},\ and\ \citenamefont {Hillebrands}}]{Chumak2015}%
  \BibitemOpen
  \bibfield  {author} {\bibinfo {author} {\bibfnamefont {A.}~\bibnamefont {Chumak}}, \bibinfo {author} {\bibfnamefont {V.}~\bibnamefont {Vasyuchka}}, \bibinfo {author} {\bibfnamefont {A.}~\bibnamefont {Serga}},\ and\ \bibinfo {author} {\bibfnamefont {B.}~\bibnamefont {Hillebrands}},\ }\bibfield  {title} {\bibinfo {title} {Magnon spintronics},\ }\href {https://doi.org/10.1038/nphys3347} {\bibfield  {journal} {\bibinfo  {journal} {Nat. Phys.}\ }\textbf {\bibinfo {volume} {11}},\ \bibinfo {pages} {453} (\bibinfo {year} {2015})}\BibitemShut {NoStop}%
\bibitem [{\citenamefont {Csaba}\ \emph {et~al.}(2017)\citenamefont {Csaba}, \citenamefont {Papp},\ and\ \citenamefont {Porod}}]{Csaba2017}%
  \BibitemOpen
  \bibfield  {author} {\bibinfo {author} {\bibfnamefont {G.}~\bibnamefont {Csaba}}, \bibinfo {author} {\bibfnamefont {A.}~\bibnamefont {Papp}},\ and\ \bibinfo {author} {\bibfnamefont {W.}~\bibnamefont {Porod}},\ }\bibfield  {title} {\bibinfo {title} {Perspectives of using spin waves for computing and signal processing},\ }\href {https://doi.org/https://doi.org/10.1016/j.physleta.2017.02.042} {\bibfield  {journal} {\bibinfo  {journal} {Phys. Lett. A}\ }\textbf {\bibinfo {volume} {381}},\ \bibinfo {pages} {1471} (\bibinfo {year} {2017})}\BibitemShut {NoStop}%
\bibitem [{\citenamefont {Chumak}()}]{Chumak2019}%
  \BibitemOpen
  \bibfield  {author} {\bibinfo {author} {\bibfnamefont {A.~V.}\ \bibnamefont {Chumak}},\ }\href@noop {} {\bibinfo {title} {Fundamentals of magnon-based computing}},\ \Eprint {https://arxiv.org/abs/1901.08934 (2019)} {arXiv:1901.08934 (2019)} \BibitemShut {NoStop}%
\bibitem [{\citenamefont {Chumak}\ \emph {et~al.}(2022)\citenamefont {Chumak}, \citenamefont {Kabos}, \citenamefont {Wu}, \citenamefont {Abert}, \citenamefont {Adelmann}, \citenamefont {Adeyeye}, \citenamefont {Akerman}, \citenamefont {Aliev}, \citenamefont {Anane}, \citenamefont {Awad} \emph {et~al.}}]{Chumak2022}%
  \BibitemOpen
  \bibfield  {author} {\bibinfo {author} {\bibfnamefont {A.~V.}\ \bibnamefont {Chumak}}, \bibinfo {author} {\bibfnamefont {P.}~\bibnamefont {Kabos}}, \bibinfo {author} {\bibfnamefont {M.}~\bibnamefont {Wu}}, \bibinfo {author} {\bibfnamefont {C.}~\bibnamefont {Abert}}, \bibinfo {author} {\bibfnamefont {C.}~\bibnamefont {Adelmann}}, \bibinfo {author} {\bibfnamefont {A.~O.}\ \bibnamefont {Adeyeye}}, \bibinfo {author} {\bibfnamefont {J.}~\bibnamefont {Akerman}}, \bibinfo {author} {\bibfnamefont {F.~G.}\ \bibnamefont {Aliev}}, \bibinfo {author} {\bibfnamefont {A.}~\bibnamefont {Anane}}, \bibinfo {author} {\bibfnamefont {A.}~\bibnamefont {Awad}}, \emph {et~al.},\ }\bibfield  {title} {\bibinfo {title} {Advances in magnetics roadmap on spin-wave computing},\ }\href {https://doi.org/10.1109/tmag.2022.3149664} {\bibfield  {journal} {\bibinfo  {journal} {{IEEE} Trans. Magn.}\ }\textbf {\bibinfo {volume} {58}},\ \bibinfo {pages} {1} (\bibinfo {year} {2022})}\BibitemShut {NoStop}%
\bibitem [{\citenamefont {Mahmoud}\ \emph {et~al.}(2020)\citenamefont {Mahmoud}, \citenamefont {Ciubotaru}, \citenamefont {Vanderveken}, \citenamefont {Chumak}, \citenamefont {Hamdioui}, \citenamefont {Adelmann},\ and\ \citenamefont {Cotofana}}]{Mahmoud2020}%
  \BibitemOpen
  \bibfield  {author} {\bibinfo {author} {\bibfnamefont {A.}~\bibnamefont {Mahmoud}}, \bibinfo {author} {\bibfnamefont {F.}~\bibnamefont {Ciubotaru}}, \bibinfo {author} {\bibfnamefont {F.}~\bibnamefont {Vanderveken}}, \bibinfo {author} {\bibfnamefont {A.~V.}\ \bibnamefont {Chumak}}, \bibinfo {author} {\bibfnamefont {S.}~\bibnamefont {Hamdioui}}, \bibinfo {author} {\bibfnamefont {C.}~\bibnamefont {Adelmann}},\ and\ \bibinfo {author} {\bibfnamefont {S.}~\bibnamefont {Cotofana}},\ }\bibfield  {title} {\bibinfo {title} {Introduction to spin wave computing},\ }\href {https://doi.org/10.1063/5.0019328} {\bibfield  {journal} {\bibinfo  {journal} {J. Appl. Phys.}\ }\textbf {\bibinfo {volume} {128}},\ \bibinfo {pages} {161101} (\bibinfo {year} {2020})}\BibitemShut {NoStop}%
\bibitem [{\citenamefont {Hamadeh}\ \emph {et~al.}(2014)\citenamefont {Hamadeh}, \citenamefont {d'Allivy Kelly}, \citenamefont {Hahn}, \citenamefont {Meley}, \citenamefont {Bernard}, \citenamefont {Molpeceres}, \citenamefont {Naletov}, \citenamefont {Viret}, \citenamefont {Anane}, \citenamefont {Cros} \emph {et~al.}}]{Hamadeh2014}%
  \BibitemOpen
  \bibfield  {author} {\bibinfo {author} {\bibfnamefont {A.}~\bibnamefont {Hamadeh}}, \bibinfo {author} {\bibfnamefont {O.}~\bibnamefont {d'Allivy Kelly}}, \bibinfo {author} {\bibfnamefont {C.}~\bibnamefont {Hahn}}, \bibinfo {author} {\bibfnamefont {H.}~\bibnamefont {Meley}}, \bibinfo {author} {\bibfnamefont {R.}~\bibnamefont {Bernard}}, \bibinfo {author} {\bibfnamefont {A.~H.}\ \bibnamefont {Molpeceres}}, \bibinfo {author} {\bibfnamefont {V.~V.}\ \bibnamefont {Naletov}}, \bibinfo {author} {\bibfnamefont {M.}~\bibnamefont {Viret}}, \bibinfo {author} {\bibfnamefont {A.}~\bibnamefont {Anane}}, \bibinfo {author} {\bibfnamefont {V.}~\bibnamefont {Cros}}, \emph {et~al.},\ }\bibfield  {title} {\bibinfo {title} {Full control of the spin-wave damping in a magnetic insulator using spin-orbit torque},\ }\href {https://doi.org/10.1103/PhysRevLett.113.197203} {\bibfield  {journal} {\bibinfo  {journal} {Phys. Rev. Lett.}\ }\textbf {\bibinfo {volume} {113}},\ \bibinfo {pages} {197203} (\bibinfo {year} {2014})}\BibitemShut
  {NoStop}%
\bibitem [{\citenamefont {Akhiezer}\ \emph {et~al.}(1964)\citenamefont {Akhiezer}, \citenamefont {Baryakhtar},\ and\ \citenamefont {Peletminskii}}]{Akhiezer1964}%
  \BibitemOpen
  \bibfield  {author} {\bibinfo {author} {\bibfnamefont {A.}~\bibnamefont {Akhiezer}}, \bibinfo {author} {\bibfnamefont {V.}~\bibnamefont {Baryakhtar}},\ and\ \bibinfo {author} {\bibfnamefont {S.}~\bibnamefont {Peletminskii}},\ }\bibfield  {title} {\bibinfo {title} {Coherent amplification of spin waves},\ }\href {https://doi.org/10.1016/0031-9163(63)90138-0} {\bibfield  {journal} {\bibinfo  {journal} {Sov. Phys. JETP}\ }\textbf {\bibinfo {volume} {18}},\ \bibinfo {pages} {235} (\bibinfo {year} {1964})}\BibitemShut {NoStop}%
\bibitem [{\citenamefont {Evelt}\ \emph {et~al.}(2016)\citenamefont {Evelt}, \citenamefont {Demidov}, \citenamefont {Bessonov}, \citenamefont {Demokritov}, \citenamefont {Prieto}, \citenamefont {Mu{\~{n}}oz}, \citenamefont {Youssef}, \citenamefont {Naletov}, \citenamefont {de~Loubens}, \citenamefont {Klein} \emph {et~al.}}]{Evelt2016}%
  \BibitemOpen
  \bibfield  {author} {\bibinfo {author} {\bibfnamefont {M.}~\bibnamefont {Evelt}}, \bibinfo {author} {\bibfnamefont {V.~E.}\ \bibnamefont {Demidov}}, \bibinfo {author} {\bibfnamefont {V.}~\bibnamefont {Bessonov}}, \bibinfo {author} {\bibfnamefont {S.~O.}\ \bibnamefont {Demokritov}}, \bibinfo {author} {\bibfnamefont {J.~L.}\ \bibnamefont {Prieto}}, \bibinfo {author} {\bibfnamefont {M.}~\bibnamefont {Mu{\~{n}}oz}}, \bibinfo {author} {\bibfnamefont {J.~B.}\ \bibnamefont {Youssef}}, \bibinfo {author} {\bibfnamefont {V.~V.}\ \bibnamefont {Naletov}}, \bibinfo {author} {\bibfnamefont {G.}~\bibnamefont {de~Loubens}}, \bibinfo {author} {\bibfnamefont {O.}~\bibnamefont {Klein}}, \emph {et~al.},\ }\bibfield  {title} {\bibinfo {title} {High-efficiency control of spin-wave propagation in ultra-thin yttrium iron garnet by the spin-orbit torque},\ }\href {https://doi.org/10.1063/1.4948252} {\bibfield  {journal} {\bibinfo  {journal} {Appl. Phys. Lett.}\ }\textbf {\bibinfo {volume} {108}},\ \bibinfo {pages} {172406}
  (\bibinfo {year} {2016})}\BibitemShut {NoStop}%
\bibitem [{\citenamefont {Demidov}\ \emph {et~al.}(2020)\citenamefont {Demidov}, \citenamefont {Urazhdin}, \citenamefont {Anane}, \citenamefont {Cros},\ and\ \citenamefont {Demokritov}}]{Demidov2020}%
  \BibitemOpen
  \bibfield  {author} {\bibinfo {author} {\bibfnamefont {V.~E.}\ \bibnamefont {Demidov}}, \bibinfo {author} {\bibfnamefont {S.}~\bibnamefont {Urazhdin}}, \bibinfo {author} {\bibfnamefont {A.}~\bibnamefont {Anane}}, \bibinfo {author} {\bibfnamefont {V.}~\bibnamefont {Cros}},\ and\ \bibinfo {author} {\bibfnamefont {S.~O.}\ \bibnamefont {Demokritov}},\ }\bibfield  {title} {\bibinfo {title} {Spin{\textendash}orbit-torque magnonics},\ }\href {https://doi.org/10.1063/5.0007095} {\bibfield  {journal} {\bibinfo  {journal} {J. Appl. Phys.}\ }\textbf {\bibinfo {volume} {127}},\ \bibinfo {pages} {170901} (\bibinfo {year} {2020})}\BibitemShut {NoStop}%
\bibitem [{\citenamefont {Breitbach}\ \emph {et~al.}(2023)\citenamefont {Breitbach}, \citenamefont {Schneider}, \citenamefont {Heinz}, \citenamefont {Kohl}, \citenamefont {Maskill}, \citenamefont {Scheuer}, \citenamefont {Serha}, \citenamefont {Br\"acher}, \citenamefont {L\"agel}, \citenamefont {Dubs} \emph {et~al.}}]{Breitbach2023}%
  \BibitemOpen
  \bibfield  {author} {\bibinfo {author} {\bibfnamefont {D.}~\bibnamefont {Breitbach}}, \bibinfo {author} {\bibfnamefont {M.}~\bibnamefont {Schneider}}, \bibinfo {author} {\bibfnamefont {B.}~\bibnamefont {Heinz}}, \bibinfo {author} {\bibfnamefont {F.}~\bibnamefont {Kohl}}, \bibinfo {author} {\bibfnamefont {J.}~\bibnamefont {Maskill}}, \bibinfo {author} {\bibfnamefont {L.}~\bibnamefont {Scheuer}}, \bibinfo {author} {\bibfnamefont {R.~O.}\ \bibnamefont {Serha}}, \bibinfo {author} {\bibfnamefont {T.}~\bibnamefont {Br\"acher}}, \bibinfo {author} {\bibfnamefont {B.}~\bibnamefont {L\"agel}}, \bibinfo {author} {\bibfnamefont {C.}~\bibnamefont {Dubs}}, \emph {et~al.},\ }\bibfield  {title} {\bibinfo {title} {Stimulated amplification of propagating spin waves},\ }\href {https://doi.org/10.1103/PhysRevLett.131.156701} {\bibfield  {journal} {\bibinfo  {journal} {Phys. Rev. Lett.}\ }\textbf {\bibinfo {volume} {131}},\ \bibinfo {pages} {156701} (\bibinfo {year} {2023})}\BibitemShut {NoStop}%
\bibitem [{\citenamefont {Bloch}(1930)}]{Bloch1930}%
  \BibitemOpen
  \bibfield  {author} {\bibinfo {author} {\bibfnamefont {F.}~\bibnamefont {Bloch}},\ }\bibfield  {title} {\bibinfo {title} {Zur theorie des ferromagnetismus},\ }\href@noop {} {\bibfield  {journal} {\bibinfo  {journal} {Z. Phys.}\ }\textbf {\bibinfo {volume} {61}},\ \bibinfo {pages} {206} (\bibinfo {year} {1930})}\BibitemShut {NoStop}%
\bibitem [{\citenamefont {Bertelli}\ \emph {et~al.}(2021)\citenamefont {Bertelli}, \citenamefont {Simon}, \citenamefont {Yu}, \citenamefont {Aarts}, \citenamefont {Bauer}, \citenamefont {Blanter},\ and\ \citenamefont {van~der Sar}}]{Bertelli2021}%
  \BibitemOpen
  \bibfield  {author} {\bibinfo {author} {\bibfnamefont {I.}~\bibnamefont {Bertelli}}, \bibinfo {author} {\bibfnamefont {B.~G.}\ \bibnamefont {Simon}}, \bibinfo {author} {\bibfnamefont {T.}~\bibnamefont {Yu}}, \bibinfo {author} {\bibfnamefont {J.}~\bibnamefont {Aarts}}, \bibinfo {author} {\bibfnamefont {G.~E.~W.}\ \bibnamefont {Bauer}}, \bibinfo {author} {\bibfnamefont {Y.~M.}\ \bibnamefont {Blanter}},\ and\ \bibinfo {author} {\bibfnamefont {T.}~\bibnamefont {van~der Sar}},\ }\bibfield  {title} {\bibinfo {title} {Imaging spin-wave damping underneath metals using electron spins in diamond},\ }\href {https://doi.org/10.1002/qute.202100094} {\bibfield  {journal} {\bibinfo  {journal} {Adv. Quantum Technol.}\ }\textbf {\bibinfo {volume} {4}},\ \bibinfo {pages} {2100094} (\bibinfo {year} {2021})}\BibitemShut {NoStop}%
\bibitem [{\citenamefont {Mae}\ \emph {et~al.}(2022)\citenamefont {Mae}, \citenamefont {Ohshima}, \citenamefont {Shigematsu}, \citenamefont {Ando}, \citenamefont {Shinjo},\ and\ \citenamefont {Shiraishi}}]{Mae2022}%
  \BibitemOpen
  \bibfield  {author} {\bibinfo {author} {\bibfnamefont {S.}~\bibnamefont {Mae}}, \bibinfo {author} {\bibfnamefont {R.}~\bibnamefont {Ohshima}}, \bibinfo {author} {\bibfnamefont {E.}~\bibnamefont {Shigematsu}}, \bibinfo {author} {\bibfnamefont {Y.}~\bibnamefont {Ando}}, \bibinfo {author} {\bibfnamefont {T.}~\bibnamefont {Shinjo}},\ and\ \bibinfo {author} {\bibfnamefont {M.}~\bibnamefont {Shiraishi}},\ }\bibfield  {title} {\bibinfo {title} {Influence of adjacent metal films on magnon propagation in {${\mathrm{Y}}_{3}{\mathrm{Fe}}_{5}{\mathrm{O}}_{12}$}},\ }\href {https://doi.org/10.1103/PhysRevB.105.104415} {\bibfield  {journal} {\bibinfo  {journal} {Phys. Rev. B}\ }\textbf {\bibinfo {volume} {105}},\ \bibinfo {pages} {104415} (\bibinfo {year} {2022})}\BibitemShut {NoStop}%
\bibitem [{\citenamefont {Krysztofik}\ \emph {et~al.}(2022)\citenamefont {Krysztofik}, \citenamefont {Kuznetsov}, \citenamefont {Qin}, \citenamefont {Flaj\v{s}man}, \citenamefont {Coy},\ and\ \citenamefont {van Dijken}}]{Krysztofik2022}%
  \BibitemOpen
  \bibfield  {author} {\bibinfo {author} {\bibfnamefont {A.}~\bibnamefont {Krysztofik}}, \bibinfo {author} {\bibfnamefont {N.}~\bibnamefont {Kuznetsov}}, \bibinfo {author} {\bibfnamefont {H.}~\bibnamefont {Qin}}, \bibinfo {author} {\bibfnamefont {L.}~\bibnamefont {Flaj\v{s}man}}, \bibinfo {author} {\bibfnamefont {E.}~\bibnamefont {Coy}},\ and\ \bibinfo {author} {\bibfnamefont {S.}~\bibnamefont {van Dijken}},\ }\bibfield  {title} {\bibinfo {title} {Tuning of magnetic damping in {Y$_3$Fe$_5$O$_{12}$}/metal bilayers for spin-wave conduit termination},\ }\href {https://doi.org/10.3390/ma15082814} {\bibfield  {journal} {\bibinfo  {journal} {Materials}\ }\textbf {\bibinfo {volume} {15}},\ \bibinfo {pages} {2814} (\bibinfo {year} {2022})}\BibitemShut {NoStop}%
\bibitem [{\citenamefont {Serha}\ \emph {et~al.}(2022)\citenamefont {Serha}, \citenamefont {Bozhko}, \citenamefont {Agrawal}, \citenamefont {Verba}, \citenamefont {Kostylev}, \citenamefont {Vasyuchka}, \citenamefont {Hillebrands},\ and\ \citenamefont {Serga}}]{Serha2022}%
  \BibitemOpen
  \bibfield  {author} {\bibinfo {author} {\bibfnamefont {R.~O.}\ \bibnamefont {Serha}}, \bibinfo {author} {\bibfnamefont {D.~A.}\ \bibnamefont {Bozhko}}, \bibinfo {author} {\bibfnamefont {M.}~\bibnamefont {Agrawal}}, \bibinfo {author} {\bibfnamefont {R.~V.}\ \bibnamefont {Verba}}, \bibinfo {author} {\bibfnamefont {M.}~\bibnamefont {Kostylev}}, \bibinfo {author} {\bibfnamefont {V.~I.}\ \bibnamefont {Vasyuchka}}, \bibinfo {author} {\bibfnamefont {B.}~\bibnamefont {Hillebrands}},\ and\ \bibinfo {author} {\bibfnamefont {A.~A.}\ \bibnamefont {Serga}},\ }\bibfield  {title} {\bibinfo {title} {Low‐damping spin‐wave transmission in {YIG}/{Pt}‐interfaced structures},\ }\href {https://doi.org/10.1002/admi.202201323} {\bibfield  {journal} {\bibinfo  {journal} {Adv. Mater. Interfaces}\ }\textbf {\bibinfo {volume} {9}},\ \bibinfo {pages} {2201323} (\bibinfo {year} {2022})}\BibitemShut {NoStop}%
\bibitem [{\citenamefont {Evans}\ \emph {et~al.}(2014)\citenamefont {Evans}, \citenamefont {Fan}, \citenamefont {Chureemart}, \citenamefont {Ostler}, \citenamefont {Ellis},\ and\ \citenamefont {Chantrell}}]{Evans2014}%
  \BibitemOpen
  \bibfield  {author} {\bibinfo {author} {\bibfnamefont {R.}~\bibnamefont {Evans}}, \bibinfo {author} {\bibfnamefont {W.}~\bibnamefont {Fan}}, \bibinfo {author} {\bibfnamefont {P.}~\bibnamefont {Chureemart}}, \bibinfo {author} {\bibfnamefont {T.}~\bibnamefont {Ostler}}, \bibinfo {author} {\bibfnamefont {M.~O.}\ \bibnamefont {Ellis}},\ and\ \bibinfo {author} {\bibfnamefont {R.}~\bibnamefont {Chantrell}},\ }\bibfield  {title} {\bibinfo {title} {Atomistic spin model simulations of magnetic nanomaterials},\ }\href {https://doi.org/10.1088/0953-8984/26/10/103202} {\bibfield  {journal} {\bibinfo  {journal} {J. Phys.: Condens. Matter}\ }\textbf {\bibinfo {volume} {26}},\ \bibinfo {pages} {103202} (\bibinfo {year} {2014})}\BibitemShut {NoStop}%
\bibitem [{\citenamefont {Kim}(2010)}]{Kim2010}%
  \BibitemOpen
  \bibfield  {author} {\bibinfo {author} {\bibfnamefont {S.-K.}\ \bibnamefont {Kim}},\ }\bibfield  {title} {\bibinfo {title} {Micromagnetic computer simulations of spin waves in nanometre-scale patterned magnetic elements},\ }\href {https://doi.org/10.1088/0022-3727/43/26/264004} {\bibfield  {journal} {\bibinfo  {journal} {J. Phys. D: Appl. Phys.}\ }\textbf {\bibinfo {volume} {43}},\ \bibinfo {pages} {264004} (\bibinfo {year} {2010})}\BibitemShut {NoStop}%
\bibitem [{\citenamefont {Serga}\ \emph {et~al.}(2010)\citenamefont {Serga}, \citenamefont {Chumak},\ and\ \citenamefont {Hillebrands}}]{Serga2010}%
  \BibitemOpen
  \bibfield  {author} {\bibinfo {author} {\bibfnamefont {A.~A.}\ \bibnamefont {Serga}}, \bibinfo {author} {\bibfnamefont {A.~V.}\ \bibnamefont {Chumak}},\ and\ \bibinfo {author} {\bibfnamefont {B.}~\bibnamefont {Hillebrands}},\ }\bibfield  {title} {\bibinfo {title} {{YIG} magnonics},\ }\href {http://stacks.iop.org/0022-3727/43/i=26/a=264002} {\bibfield  {journal} {\bibinfo  {journal} {J. Phys. D: Appl. Phys.}\ }\textbf {\bibinfo {volume} {43}},\ \bibinfo {pages} {264002} (\bibinfo {year} {2010})}\BibitemShut {NoStop}%
\bibitem [{\citenamefont {Bihlmayer}\ \emph {et~al.}(2022)\citenamefont {Bihlmayer}, \citenamefont {Noël}, \citenamefont {Vyalikh}, \citenamefont {Chulkov},\ and\ \citenamefont {Manchon}}]{Bihlmayer2022}%
  \BibitemOpen
  \bibfield  {author} {\bibinfo {author} {\bibfnamefont {G.}~\bibnamefont {Bihlmayer}}, \bibinfo {author} {\bibfnamefont {P.}~\bibnamefont {Noël}}, \bibinfo {author} {\bibfnamefont {D.~V.}\ \bibnamefont {Vyalikh}}, \bibinfo {author} {\bibfnamefont {E.~V.}\ \bibnamefont {Chulkov}},\ and\ \bibinfo {author} {\bibfnamefont {A.}~\bibnamefont {Manchon}},\ }\bibfield  {title} {\bibinfo {title} {Rashba-like physics in condensed matter},\ }\href {https://doi.org/10.1038/s42254-022-00490-y} {\bibfield  {journal} {\bibinfo  {journal} {Nat. Rev. Phys.}\ }\textbf {\bibinfo {volume} {4}},\ \bibinfo {pages} {642} (\bibinfo {year} {2022})}\BibitemShut {NoStop}%
\bibitem [{\citenamefont {Manchon}\ \emph {et~al.}(2015)\citenamefont {Manchon}, \citenamefont {Koo}, \citenamefont {Nitta}, \citenamefont {Frolov},\ and\ \citenamefont {Duine}}]{Manchon2015}%
  \BibitemOpen
  \bibfield  {author} {\bibinfo {author} {\bibfnamefont {A.}~\bibnamefont {Manchon}}, \bibinfo {author} {\bibfnamefont {H.~C.}\ \bibnamefont {Koo}}, \bibinfo {author} {\bibfnamefont {J.}~\bibnamefont {Nitta}}, \bibinfo {author} {\bibfnamefont {S.~M.}\ \bibnamefont {Frolov}},\ and\ \bibinfo {author} {\bibfnamefont {R.~A.}\ \bibnamefont {Duine}},\ }\bibfield  {title} {\bibinfo {title} {New perspectives for {Rashba} spin{\textendash}orbit coupling},\ }\href {https://doi.org/10.1038/nmat4360} {\bibfield  {journal} {\bibinfo  {journal} {Nat. Mater.}\ }\textbf {\bibinfo {volume} {14}},\ \bibinfo {pages} {871} (\bibinfo {year} {2015})}\BibitemShut {NoStop}%
\bibitem [{\citenamefont {\ifmmode~\check{S}\else \v{S}\fi{}mejkal}\ \emph {et~al.}(2022{\natexlab{a}})\citenamefont {\ifmmode~\check{S}\else \v{S}\fi{}mejkal}, \citenamefont {Sinova},\ and\ \citenamefont {Jungwirth}}]{Smejkal2022b}%
  \BibitemOpen
  \bibfield  {author} {\bibinfo {author} {\bibfnamefont {L.}~\bibnamefont {\ifmmode~\check{S}\else \v{S}\fi{}mejkal}}, \bibinfo {author} {\bibfnamefont {J.}~\bibnamefont {Sinova}},\ and\ \bibinfo {author} {\bibfnamefont {T.}~\bibnamefont {Jungwirth}},\ }\bibfield  {title} {\bibinfo {title} {Emerging research landscape of altermagnetism},\ }\href {https://doi.org/10.1103/PhysRevX.12.040501} {\bibfield  {journal} {\bibinfo  {journal} {Phys. Rev. X}\ }\textbf {\bibinfo {volume} {12}},\ \bibinfo {pages} {040501} (\bibinfo {year} {2022}{\natexlab{a}})}\BibitemShut {NoStop}%
\bibitem [{\citenamefont {\ifmmode~\check{S}\else \v{S}\fi{}mejkal}\ \emph {et~al.}(2022{\natexlab{b}})\citenamefont {\ifmmode~\check{S}\else \v{S}\fi{}mejkal}, \citenamefont {Hellenes}, \citenamefont {Gonz\'alez-Hern\'andez}, \citenamefont {Sinova},\ and\ \citenamefont {Jungwirth}}]{Smejkal2022a}%
  \BibitemOpen
  \bibfield  {author} {\bibinfo {author} {\bibfnamefont {L.}~\bibnamefont {\ifmmode~\check{S}\else \v{S}\fi{}mejkal}}, \bibinfo {author} {\bibfnamefont {A.~B.}\ \bibnamefont {Hellenes}}, \bibinfo {author} {\bibfnamefont {R.}~\bibnamefont {Gonz\'alez-Hern\'andez}}, \bibinfo {author} {\bibfnamefont {J.}~\bibnamefont {Sinova}},\ and\ \bibinfo {author} {\bibfnamefont {T.}~\bibnamefont {Jungwirth}},\ }\bibfield  {title} {\bibinfo {title} {Giant and tunneling magnetoresistance in unconventional collinear antiferromagnets with nonrelativistic spin-momentum coupling},\ }\href {https://doi.org/10.1103/PhysRevX.12.011028} {\bibfield  {journal} {\bibinfo  {journal} {Phys. Rev. X}\ }\textbf {\bibinfo {volume} {12}},\ \bibinfo {pages} {011028} (\bibinfo {year} {2022}{\natexlab{b}})}\BibitemShut {NoStop}%
\bibitem [{\citenamefont {Gilbert}(2004)}]{Gilbert2004}%
  \BibitemOpen
  \bibfield  {author} {\bibinfo {author} {\bibfnamefont {T.}~\bibnamefont {Gilbert}},\ }\bibfield  {title} {\bibinfo {title} {A phenomenological theory of damping in ferromagnetic materials},\ }\href {https://doi.org/10.1109/TMAG.2004.836740} {\bibfield  {journal} {\bibinfo  {journal} {{IEEE} Trans. Magn.}\ }\textbf {\bibinfo {volume} {40}},\ \bibinfo {pages} {3443} (\bibinfo {year} {2004})}\BibitemShut {NoStop}%
\bibitem [{\citenamefont {Saslow}(2009)}]{Saslow2009}%
  \BibitemOpen
  \bibfield  {author} {\bibinfo {author} {\bibfnamefont {W.~M.}\ \bibnamefont {Saslow}},\ }\bibfield  {title} {\bibinfo {title} {{L}andau-{L}ifshitz or {G}ilbert damping? that is the question},\ }\href {https://doi.org/10.1063/1.3077204} {\bibfield  {journal} {\bibinfo  {journal} {J. Appl. Phys.}\ }\textbf {\bibinfo {volume} {105}},\ \bibinfo {pages} {07D315} (\bibinfo {year} {2009})}\BibitemShut {NoStop}%
\bibitem [{\citenamefont {Zhang}\ and\ \citenamefont {Zhang}(2009)}]{Zhang2009}%
  \BibitemOpen
  \bibfield  {author} {\bibinfo {author} {\bibfnamefont {S.}~\bibnamefont {Zhang}}\ and\ \bibinfo {author} {\bibfnamefont {S.~L.}\ \bibnamefont {Zhang}},\ }\bibfield  {title} {\bibinfo {title} {Generalization of the {Landau}-{Lifshitz}-{Gilbert} equation for conducting ferromagnets.},\ }\href {https://doi.org/10.1103/PHYSREVLETT.102.086601} {\bibfield  {journal} {\bibinfo  {journal} {Phys. Rev. Lett.}\ }\textbf {\bibinfo {volume} {102}},\ \bibinfo {pages} {086601} (\bibinfo {year} {2009})}\BibitemShut {NoStop}%
\bibitem [{\citenamefont {Yuan}\ \emph {et~al.}(2016)\citenamefont {Yuan}, \citenamefont {Yuan}, \citenamefont {Xia},\ and\ \citenamefont {Wang}}]{Yuan2016}%
  \BibitemOpen
  \bibfield  {author} {\bibinfo {author} {\bibfnamefont {H.}~\bibnamefont {Yuan}}, \bibinfo {author} {\bibfnamefont {Z.}~\bibnamefont {Yuan}}, \bibinfo {author} {\bibfnamefont {K.}~\bibnamefont {Xia}},\ and\ \bibinfo {author} {\bibfnamefont {X.~R.}\ \bibnamefont {Wang}},\ }\bibfield  {title} {\bibinfo {title} {Influence of nonlocal damping on the field-driven domain wall motion},\ }\href {https://doi.org/10.1103/PhysRevB.94.064415} {\bibfield  {journal} {\bibinfo  {journal} {Phys. Rev. B}\ }\textbf {\bibinfo {volume} {94}},\ \bibinfo {pages} {064415} (\bibinfo {year} {2016})}\BibitemShut {NoStop}%
\bibitem [{\citenamefont {Verba}\ \emph {et~al.}(2018)\citenamefont {Verba}, \citenamefont {Tiberkevich},\ and\ \citenamefont {Slavin}}]{Verba2018}%
  \BibitemOpen
  \bibfield  {author} {\bibinfo {author} {\bibfnamefont {R.}~\bibnamefont {Verba}}, \bibinfo {author} {\bibfnamefont {V.}~\bibnamefont {Tiberkevich}},\ and\ \bibinfo {author} {\bibfnamefont {A.}~\bibnamefont {Slavin}},\ }\bibfield  {title} {\bibinfo {title} {Damping of linear spin-wave modes in magnetic nanostructures: Local, nonlocal, and coordinate-dependent damping},\ }\href {https://doi.org/10.1103/PHYSREVB.98.104408} {\bibfield  {journal} {\bibinfo  {journal} {Phys. Rev. B}\ }\textbf {\bibinfo {volume} {98}},\ \bibinfo {pages} {104408} (\bibinfo {year} {2018})}\BibitemShut {NoStop}%
\bibitem [{\citenamefont {Lu}\ \emph {et~al.}(2023)\citenamefont {Lu}, \citenamefont {Miranda}, \citenamefont {Streib}, \citenamefont {Pereiro}, \citenamefont {Sj\"oqvist}, \citenamefont {Eriksson}, \citenamefont {Bergman}, \citenamefont {Thonig},\ and\ \citenamefont {Delin}}]{Lu2023}%
  \BibitemOpen
  \bibfield  {author} {\bibinfo {author} {\bibfnamefont {Z.}~\bibnamefont {Lu}}, \bibinfo {author} {\bibfnamefont {I.~P.}\ \bibnamefont {Miranda}}, \bibinfo {author} {\bibfnamefont {S.}~\bibnamefont {Streib}}, \bibinfo {author} {\bibfnamefont {M.}~\bibnamefont {Pereiro}}, \bibinfo {author} {\bibfnamefont {E.}~\bibnamefont {Sj\"oqvist}}, \bibinfo {author} {\bibfnamefont {O.}~\bibnamefont {Eriksson}}, \bibinfo {author} {\bibfnamefont {A.}~\bibnamefont {Bergman}}, \bibinfo {author} {\bibfnamefont {D.}~\bibnamefont {Thonig}},\ and\ \bibinfo {author} {\bibfnamefont {A.}~\bibnamefont {Delin}},\ }\bibfield  {title} {\bibinfo {title} {Influence of nonlocal damping on magnon properties of ferromagnets},\ }\href {https://doi.org/10.1103/PhysRevB.108.014433} {\bibfield  {journal} {\bibinfo  {journal} {Phys. Rev. B}\ }\textbf {\bibinfo {volume} {108}},\ \bibinfo {pages} {014433} (\bibinfo {year} {2023})}\BibitemShut {NoStop}%
\bibitem [{\citenamefont {Gilmore}\ \emph {et~al.}(2007)\citenamefont {Gilmore}, \citenamefont {Idzerda},\ and\ \citenamefont {Stiles}}]{Gilmore2007}%
  \BibitemOpen
  \bibfield  {author} {\bibinfo {author} {\bibfnamefont {K.}~\bibnamefont {Gilmore}}, \bibinfo {author} {\bibfnamefont {Y.}~\bibnamefont {Idzerda}},\ and\ \bibinfo {author} {\bibfnamefont {M.}~\bibnamefont {Stiles}},\ }\bibfield  {title} {\bibinfo {title} {Identification of the dominant precession-damping mechanism in {Fe}, {Co}, and {Ni} by first-principles calculations.},\ }\href {https://doi.org/10.1103/PhysRevLett.99.027204} {\bibfield  {journal} {\bibinfo  {journal} {Phys. Rev. Lett.}\ }\textbf {\bibinfo {volume} {99}},\ \bibinfo {pages} {027204} (\bibinfo {year} {2007})}\BibitemShut {NoStop}%
\bibitem [{\citenamefont {Trempler}\ \emph {et~al.}(2020)\citenamefont {Trempler}, \citenamefont {Dreyer}, \citenamefont {Geyer}, \citenamefont {Hauser}, \citenamefont {Woltersdorf},\ and\ \citenamefont {Schmidt}}]{Trempler2020}%
  \BibitemOpen
  \bibfield  {author} {\bibinfo {author} {\bibfnamefont {P.}~\bibnamefont {Trempler}}, \bibinfo {author} {\bibfnamefont {R.}~\bibnamefont {Dreyer}}, \bibinfo {author} {\bibfnamefont {P.}~\bibnamefont {Geyer}}, \bibinfo {author} {\bibfnamefont {C.}~\bibnamefont {Hauser}}, \bibinfo {author} {\bibfnamefont {G.}~\bibnamefont {Woltersdorf}},\ and\ \bibinfo {author} {\bibfnamefont {G.}~\bibnamefont {Schmidt}},\ }\bibfield  {title} {\bibinfo {title} {Integration and characterization of micron-sized {YIG} structures with very low {Gilbert} damping on arbitrary substrates},\ }\href {https://doi.org/10.1063/5.0026120} {\bibfield  {journal} {\bibinfo  {journal} {Appl. Phys. Lett.}\ }\textbf {\bibinfo {volume} {117}},\ \bibinfo {pages} {232401} (\bibinfo {year} {2020})}\BibitemShut {NoStop}%
\bibitem [{\citenamefont {Casola}\ \emph {et~al.}(2018)\citenamefont {Casola}, \citenamefont {van~der Sar},\ and\ \citenamefont {Yacoby}}]{Casola2018}%
  \BibitemOpen
  \bibfield  {author} {\bibinfo {author} {\bibfnamefont {F.}~\bibnamefont {Casola}}, \bibinfo {author} {\bibfnamefont {T.}~\bibnamefont {van~der Sar}},\ and\ \bibinfo {author} {\bibfnamefont {A.}~\bibnamefont {Yacoby}},\ }\bibfield  {title} {\bibinfo {title} {Probing condensed matter physics with magnetometry based on nitrogen-vacancy centres in diamond},\ }\href {https://doi.org/10.1038/natrevmats.2017.88} {\bibfield  {journal} {\bibinfo  {journal} {Nat. Rev. Mater.}\ }\textbf {\bibinfo {volume} {3}},\ \bibinfo {pages} {17088} (\bibinfo {year} {2018})}\BibitemShut {NoStop}%
\bibitem [{\citenamefont {Krivorotov}\ \emph {et~al.}(2007)\citenamefont {Krivorotov}, \citenamefont {Berkov}, \citenamefont {Gorn}, \citenamefont {Emley}, \citenamefont {Sankey}, \citenamefont {Ralph},\ and\ \citenamefont {Buhrman}}]{Krivorotov2007}%
  \BibitemOpen
  \bibfield  {author} {\bibinfo {author} {\bibfnamefont {I.~N.}\ \bibnamefont {Krivorotov}}, \bibinfo {author} {\bibfnamefont {D.~V.}\ \bibnamefont {Berkov}}, \bibinfo {author} {\bibfnamefont {N.~L.}\ \bibnamefont {Gorn}}, \bibinfo {author} {\bibfnamefont {N.~C.}\ \bibnamefont {Emley}}, \bibinfo {author} {\bibfnamefont {J.~C.}\ \bibnamefont {Sankey}}, \bibinfo {author} {\bibfnamefont {D.~C.}\ \bibnamefont {Ralph}},\ and\ \bibinfo {author} {\bibfnamefont {R.~A.}\ \bibnamefont {Buhrman}},\ }\bibfield  {title} {\bibinfo {title} {Large-amplitude coherent spin waves excited by spin-polarized current in nanoscale spin valves},\ }\href {https://doi.org/10.1103/PhysRevB.76.024418} {\bibfield  {journal} {\bibinfo  {journal} {Phys. Rev. B}\ }\textbf {\bibinfo {volume} {76}},\ \bibinfo {pages} {024418} (\bibinfo {year} {2007})}\BibitemShut {NoStop}%
\bibitem [{\citenamefont {Kapelrud}\ and\ \citenamefont {Brataas}(2013)}]{Kapelrud2013}%
  \BibitemOpen
  \bibfield  {author} {\bibinfo {author} {\bibfnamefont {A.}~\bibnamefont {Kapelrud}}\ and\ \bibinfo {author} {\bibfnamefont {A.}~\bibnamefont {Brataas}},\ }\bibfield  {title} {\bibinfo {title} {Spin pumping and enhanced {Gilbert} damping in thin magnetic insulator films},\ }\href {https://doi.org/10.1103/PhysRevLett.111.097602} {\bibfield  {journal} {\bibinfo  {journal} {Phys. Rev. Lett.}\ }\textbf {\bibinfo {volume} {111}},\ \bibinfo {pages} {097602} (\bibinfo {year} {2013})}\BibitemShut {NoStop}%
\bibitem [{\citenamefont {Skarsv\aa{}g}\ \emph {et~al.}(2014)\citenamefont {Skarsv\aa{}g}, \citenamefont {Kapelrud},\ and\ \citenamefont {Brataas}}]{Skarsvaag2014}%
  \BibitemOpen
  \bibfield  {author} {\bibinfo {author} {\bibfnamefont {H.}~\bibnamefont {Skarsv\aa{}g}}, \bibinfo {author} {\bibfnamefont {A.}~\bibnamefont {Kapelrud}},\ and\ \bibinfo {author} {\bibfnamefont {A.}~\bibnamefont {Brataas}},\ }\bibfield  {title} {\bibinfo {title} {Spin waves in ferromagnetic insulators coupled via a normal metal},\ }\href {https://doi.org/10.1103/PhysRevB.90.094418} {\bibfield  {journal} {\bibinfo  {journal} {Phys. Rev. B}\ }\textbf {\bibinfo {volume} {90}},\ \bibinfo {pages} {094418} (\bibinfo {year} {2014})}\BibitemShut {NoStop}%
\bibitem [{\citenamefont {Reyes-Osorio}\ and\ \citenamefont {Nikoli\ifmmode~\acute{c}\else \'{c}\fi{}}(2024)}]{ReyesOsorio2023a}%
  \BibitemOpen
  \bibfield  {author} {\bibinfo {author} {\bibfnamefont {F.}~\bibnamefont {Reyes-Osorio}}\ and\ \bibinfo {author} {\bibfnamefont {B.~K.}\ \bibnamefont {Nikoli\ifmmode~\acute{c}\else \'{c}\fi{}}},\ }\bibfield  {title} {\bibinfo {title} {Gilbert damping in metallic ferromagnets from {Schwinger-Keldysh} field theory: Intrinsically nonlocal, nonuniform, and made anisotropic by spin-orbit coupling},\ }\href {https://doi.org/10.1103/PhysRevB.109.024413} {\bibfield  {journal} {\bibinfo  {journal} {Phys. Rev. B}\ }\textbf {\bibinfo {volume} {109}},\ \bibinfo {pages} {024413} (\bibinfo {year} {2024})}\BibitemShut {NoStop}%
\bibitem [{\citenamefont {Kamenev}(2023)}]{Kamenev2011}%
  \BibitemOpen
  \bibfield  {author} {\bibinfo {author} {\bibfnamefont {A.}~\bibnamefont {Kamenev}},\ }\href {https://doi.org/10.1017/CBO9781139003667} {\emph {\bibinfo {title} {Field Theory of Non-Equilibrium Systems}}}\ (\bibinfo  {publisher} {Cambridge University Press, Cambridge},\ \bibinfo {year} {2023})\BibitemShut {NoStop}%
\bibitem [{\citenamefont {Petrovi\'c}\ \emph {et~al.}(2018)\citenamefont {Petrovi\'c}, \citenamefont {Popescu}, \citenamefont {Plech\'a{\v c}},\ and\ \citenamefont {Nikoli\'c}}]{Petrovic2018}%
  \BibitemOpen
  \bibfield  {author} {\bibinfo {author} {\bibfnamefont {M.}~\bibnamefont {Petrovi\'c}}, \bibinfo {author} {\bibfnamefont {B.}~\bibnamefont {Popescu}}, \bibinfo {author} {\bibfnamefont {P.}~\bibnamefont {Plech\'a{\v c}}},\ and\ \bibinfo {author} {\bibfnamefont {B.}~\bibnamefont {Nikoli\'c}},\ }\bibfield  {title} {\bibinfo {title} {Spin and charge pumping by current-driven magnetic domain wall motion: A self-consistent multiscale time-dependent-quantum/time-dependent-classical approach},\ }\href {https://doi.org/10.1103/PhysRevApplied.10.054038} {\bibfield  {journal} {\bibinfo  {journal} {Phys. Rev. Appl.}\ }\textbf {\bibinfo {volume} {10}},\ \bibinfo {pages} {054038} (\bibinfo {year} {2018})}\BibitemShut {NoStop}%
\bibitem [{\citenamefont {Bajpai}\ and\ \citenamefont {Nikoli\'c}(2019)}]{Bajpai2019}%
  \BibitemOpen
  \bibfield  {author} {\bibinfo {author} {\bibfnamefont {U.}~\bibnamefont {Bajpai}}\ and\ \bibinfo {author} {\bibfnamefont {B.}~\bibnamefont {Nikoli\'c}},\ }\bibfield  {title} {\bibinfo {title} {Time-retarded damping and magnetic inertia in the {Landau}-{Lifshitz}-{Gilbert} equation self-consistently coupled to electronic time-dependent nonequilibrium {Green} functions},\ }\href {https://doi.org/10.1103/PhysRevB.99.134409} {\bibfield  {journal} {\bibinfo  {journal} {Phys. Rev. B}\ }\textbf {\bibinfo {volume} {99}},\ \bibinfo {pages} {134409} (\bibinfo {year} {2019})}\BibitemShut {NoStop}%
\bibitem [{\citenamefont {Petrovi{\'{c}}}\ \emph {et~al.}(2021)\citenamefont {Petrovi{\'{c}}}, \citenamefont {Bajpai}, \citenamefont {Plech{\'{a}}{\v{c}}},\ and\ \citenamefont {Nikoli{\'{c}}}}]{Petrovic2021}%
  \BibitemOpen
  \bibfield  {author} {\bibinfo {author} {\bibfnamefont {M.~D.}\ \bibnamefont {Petrovi{\'{c}}}}, \bibinfo {author} {\bibfnamefont {U.}~\bibnamefont {Bajpai}}, \bibinfo {author} {\bibfnamefont {P.}~\bibnamefont {Plech{\'{a}}{\v{c}}}},\ and\ \bibinfo {author} {\bibfnamefont {B.~K.}\ \bibnamefont {Nikoli{\'{c}}}},\ }\bibfield  {title} {\bibinfo {title} {Annihilation of topological solitons in magnetism with spin-wave burst finale: Role of nonequilibrium electrons causing nonlocal damping and spin pumping over ultrabroadband frequency range},\ }\href {https://doi.org/10.1103/PhysRevB.104.L020407} {\bibfield  {journal} {\bibinfo  {journal} {Phys. Rev. B}\ }\textbf {\bibinfo {volume} {104}},\ \bibinfo {pages} {l020407} (\bibinfo {year} {2021})}\BibitemShut {NoStop}%
\bibitem [{\citenamefont {Suresh}\ \emph {et~al.}(2020)\citenamefont {Suresh}, \citenamefont {Bajpai},\ and\ \citenamefont {Nikoli{\'c}}}]{Suresh2020}%
  \BibitemOpen
  \bibfield  {author} {\bibinfo {author} {\bibfnamefont {A.}~\bibnamefont {Suresh}}, \bibinfo {author} {\bibfnamefont {U.}~\bibnamefont {Bajpai}},\ and\ \bibinfo {author} {\bibfnamefont {B.~K.}\ \bibnamefont {Nikoli{\'c}}},\ }\bibfield  {title} {\bibinfo {title} {Magnon-driven chiral charge and spin pumping and electron-magnon scattering from time-dependent quantum transport combined with classical atomistic spin dynamics},\ }\href {https://doi.org/10.1103/PhysRevB.101.214412} {\bibfield  {journal} {\bibinfo  {journal} {Phys. Rev. B}\ }\textbf {\bibinfo {volume} {101}},\ \bibinfo {pages} {214412} (\bibinfo {year} {2020})}\BibitemShut {NoStop}%
\bibitem [{\citenamefont {Haehl}\ \emph {et~al.}(2017{\natexlab{a}})\citenamefont {Haehl}, \citenamefont {Loganayagam},\ and\ \citenamefont {Rangamani}}]{Haehl2017}%
  \BibitemOpen
  \bibfield  {author} {\bibinfo {author} {\bibfnamefont {F.~M.}\ \bibnamefont {Haehl}}, \bibinfo {author} {\bibfnamefont {R.}~\bibnamefont {Loganayagam}},\ and\ \bibinfo {author} {\bibfnamefont {M.}~\bibnamefont {Rangamani}},\ }\bibfield  {title} {\bibinfo {title} {Schwinger-{Keldysh} formalism. {Part I}: {BRST} symmetries and superspace},\ }\href {https://doi.org/10.1007/JHEP06(2017)069} {\bibfield  {journal} {\bibinfo  {journal} {J. High Energy Phys{.}}\ }\textbf {\bibinfo {volume} {2017}},\ \bibinfo {pages} {69} (\bibinfo {year} {2017}{\natexlab{a}})}\BibitemShut {NoStop}%
\bibitem [{\citenamefont {Haehl}\ \emph {et~al.}(2017{\natexlab{b}})\citenamefont {Haehl}, \citenamefont {Loganayagam},\ and\ \citenamefont {Rangamani}}]{Haehl2017a}%
  \BibitemOpen
  \bibfield  {author} {\bibinfo {author} {\bibfnamefont {F.~M.}\ \bibnamefont {Haehl}}, \bibinfo {author} {\bibfnamefont {R.}~\bibnamefont {Loganayagam}},\ and\ \bibinfo {author} {\bibfnamefont {M.}~\bibnamefont {Rangamani}},\ }\bibfield  {title} {\bibinfo {title} {{Schwinger}-{Keldysh} formalism. {Part} {II}: thermal equivariant cohomology},\ }\href {https://doi.org/10.1007/JHEP06(2017)070} {\bibfield  {journal} {\bibinfo  {journal} {J. High Energy Phys{.}}\ }\textbf {\bibinfo {volume} {2017}},\ \bibinfo {pages} {70} (\bibinfo {year} {2017}{\natexlab{b}})}\BibitemShut {NoStop}%
\bibitem [{\citenamefont {Berges}()}]{Berges2015}%
  \BibitemOpen
  \bibfield  {author} {\bibinfo {author} {\bibfnamefont {J.}~\bibnamefont {Berges}},\ }\href@noop {} {\bibinfo {title} {Nonequilibrium quantum fields: From cold atoms to cosmology}},\ \Eprint {https://arxiv.org/abs/1503.02907 (2015)} {arXiv:1503.02907 (2015)} \BibitemShut {NoStop}%
\bibitem [{\citenamefont {Anders}\ \emph {et~al.}(2022)\citenamefont {Anders}, \citenamefont {Sait},\ and\ \citenamefont {Horsley}}]{Anders2022}%
  \BibitemOpen
  \bibfield  {author} {\bibinfo {author} {\bibfnamefont {J.}~\bibnamefont {Anders}}, \bibinfo {author} {\bibfnamefont {C.~R.~J.}\ \bibnamefont {Sait}},\ and\ \bibinfo {author} {\bibfnamefont {S.~A.~R.}\ \bibnamefont {Horsley}},\ }\bibfield  {title} {\bibinfo {title} {Quantum {Brownian} motion for magnets},\ }\href {https://doi.org/10.1088/1367-2630/ac4ef2} {\bibfield  {journal} {\bibinfo  {journal} {New J. Phys.}\ }\textbf {\bibinfo {volume} {24}},\ \bibinfo {pages} {033020} (\bibinfo {year} {2022})}\BibitemShut {NoStop}%
\bibitem [{\citenamefont {Leiva~M.}\ \emph {et~al.}(2023)\citenamefont {Leiva~M.}, \citenamefont {D\'{\i}az},\ and\ \citenamefont {Nunez}}]{Leiva2023}%
  \BibitemOpen
  \bibfield  {author} {\bibinfo {author} {\bibfnamefont {S.}~\bibnamefont {Leiva~M.}}, \bibinfo {author} {\bibfnamefont {S.~A.}\ \bibnamefont {D\'{\i}az}},\ and\ \bibinfo {author} {\bibfnamefont {A.~S.}\ \bibnamefont {Nunez}},\ }\bibfield  {title} {\bibinfo {title} {Origin of the magnetoelectric couplings in the spin dynamics of molecular magnets},\ }\href {https://doi.org/10.1103/PhysRevB.107.094401} {\bibfield  {journal} {\bibinfo  {journal} {Phys. Rev. B}\ }\textbf {\bibinfo {volume} {107}},\ \bibinfo {pages} {094401} (\bibinfo {year} {2023})}\BibitemShut {NoStop}%
\bibitem [{\citenamefont {Bai}\ \emph {et~al.}(2023)\citenamefont {Bai}, \citenamefont {Zhang}, \citenamefont {Zhou}, \citenamefont {Chen}, \citenamefont {Wan}, \citenamefont {Han}, \citenamefont {Zhu}, \citenamefont {Liang}, \citenamefont {Su}, \citenamefont {Han} \emph {et~al.}}]{Bai2023}%
  \BibitemOpen
  \bibfield  {author} {\bibinfo {author} {\bibfnamefont {H.}~\bibnamefont {Bai}}, \bibinfo {author} {\bibfnamefont {Y.~C.}\ \bibnamefont {Zhang}}, \bibinfo {author} {\bibfnamefont {Y.~J.}\ \bibnamefont {Zhou}}, \bibinfo {author} {\bibfnamefont {P.}~\bibnamefont {Chen}}, \bibinfo {author} {\bibfnamefont {C.~H.}\ \bibnamefont {Wan}}, \bibinfo {author} {\bibfnamefont {L.}~\bibnamefont {Han}}, \bibinfo {author} {\bibfnamefont {W.~X.}\ \bibnamefont {Zhu}}, \bibinfo {author} {\bibfnamefont {S.~X.}\ \bibnamefont {Liang}}, \bibinfo {author} {\bibfnamefont {Y.~C.}\ \bibnamefont {Su}}, \bibinfo {author} {\bibfnamefont {X.~F.}\ \bibnamefont {Han}}, \emph {et~al.},\ }\bibfield  {title} {\bibinfo {title} {Efficient spin-to-charge conversion via altermagnetic spin splitting effect in antiferromagnet {${\mathrm{RuO}}_{2}$}},\ }\href {https://doi.org/10.1103/PhysRevLett.130.216701} {\bibfield  {journal} {\bibinfo  {journal} {Phys. Rev. Lett.}\ }\textbf {\bibinfo {volume} {130}},\ \bibinfo {pages} {216701} (\bibinfo {year}
  {2023})}\BibitemShut {NoStop}%
\bibitem [{\citenamefont {Madami}\ \emph {et~al.}(2011)\citenamefont {Madami}, \citenamefont {Bonetti}, \citenamefont {Consolo}, \citenamefont {Tacchi}, \citenamefont {Carlotti}, \citenamefont {Gubbiotti}, \citenamefont {Mancoff}, \citenamefont {Yar},\ and\ \citenamefont {{\AA}kerman}}]{Madami2011}%
  \BibitemOpen
  \bibfield  {author} {\bibinfo {author} {\bibfnamefont {M.}~\bibnamefont {Madami}}, \bibinfo {author} {\bibfnamefont {S.}~\bibnamefont {Bonetti}}, \bibinfo {author} {\bibfnamefont {G.}~\bibnamefont {Consolo}}, \bibinfo {author} {\bibfnamefont {S.}~\bibnamefont {Tacchi}}, \bibinfo {author} {\bibfnamefont {G.}~\bibnamefont {Carlotti}}, \bibinfo {author} {\bibfnamefont {G.}~\bibnamefont {Gubbiotti}}, \bibinfo {author} {\bibfnamefont {F.~B.}\ \bibnamefont {Mancoff}}, \bibinfo {author} {\bibfnamefont {M.~A.}\ \bibnamefont {Yar}},\ and\ \bibinfo {author} {\bibfnamefont {J.}~\bibnamefont {{\AA}kerman}},\ }\bibfield  {title} {\bibinfo {title} {Direct observation of a propagating spin wave induced by spin-transfer torque},\ }\href {https://doi.org/10.1038/nnano.2011.140} {\bibfield  {journal} {\bibinfo  {journal} {Nat. Nanotechnol.}\ }\textbf {\bibinfo {volume} {6}},\ \bibinfo {pages} {635} (\bibinfo {year} {2011})}\BibitemShut {NoStop}%
\bibitem [{\citenamefont {Halperin}\ and\ \citenamefont {Hohenberg}(1969)}]{Halperin1969}%
  \BibitemOpen
  \bibfield  {author} {\bibinfo {author} {\bibfnamefont {B.}~\bibnamefont {Halperin}}\ and\ \bibinfo {author} {\bibfnamefont {P.}~\bibnamefont {Hohenberg}},\ }\bibfield  {title} {\bibinfo {title} {Hydrodynamic theory of spin waves},\ }\href {https://doi.org/10.1103/PHYSREV.188.898} {\bibfield  {journal} {\bibinfo  {journal} {Phys. Rev.}\ }\textbf {\bibinfo {volume} {188}},\ \bibinfo {pages} {898} (\bibinfo {year} {1969})}\BibitemShut {NoStop}%
\bibitem [{\citenamefont {Sayad}\ and\ \citenamefont {Potthoff}(2015)}]{Sayad2015}%
  \BibitemOpen
  \bibfield  {author} {\bibinfo {author} {\bibfnamefont {M.}~\bibnamefont {Sayad}}\ and\ \bibinfo {author} {\bibfnamefont {M.}~\bibnamefont {Potthoff}},\ }\bibfield  {title} {\bibinfo {title} {Spin dynamics and relaxation in the classical-spin {Kondo}-impurity model beyond the {Landau}-{Lifschitz}-{Gilbert} equation},\ }\href {https://doi.org/10.1088/1367-2630/17/11/113058} {\bibfield  {journal} {\bibinfo  {journal} {New J. Phys.}\ }\textbf {\bibinfo {volume} {17}},\ \bibinfo {pages} {113058} (\bibinfo {year} {2015})}\BibitemShut {NoStop}%
\bibitem [{\citenamefont {Ralph}\ and\ \citenamefont {Stiles}(2008)}]{Ralph2008}%
  \BibitemOpen
  \bibfield  {author} {\bibinfo {author} {\bibfnamefont {D.}~\bibnamefont {Ralph}}\ and\ \bibinfo {author} {\bibfnamefont {M.}~\bibnamefont {Stiles}},\ }\bibfield  {title} {\bibinfo {title} {Spin transfer torques},\ }\href {https://doi.org/10.1016/j.jmmm.2007.12.019} {\bibfield  {journal} {\bibinfo  {journal} {J. Magn. Magn. Mater.}\ }\textbf {\bibinfo {volume} {320}},\ \bibinfo {pages} {1190} (\bibinfo {year} {2008})}\BibitemShut {NoStop}%
\bibitem [{\citenamefont {Gaury}\ \emph {et~al.}(2014)\citenamefont {Gaury}, \citenamefont {Weston}, \citenamefont {Santin}, \citenamefont {Houzet}, \citenamefont {Groth},\ and\ \citenamefont {Waintal}}]{Gaury2014}%
  \BibitemOpen
  \bibfield  {author} {\bibinfo {author} {\bibfnamefont {B.}~\bibnamefont {Gaury}}, \bibinfo {author} {\bibfnamefont {J.}~\bibnamefont {Weston}}, \bibinfo {author} {\bibfnamefont {M.}~\bibnamefont {Santin}}, \bibinfo {author} {\bibfnamefont {M.}~\bibnamefont {Houzet}}, \bibinfo {author} {\bibfnamefont {C.}~\bibnamefont {Groth}},\ and\ \bibinfo {author} {\bibfnamefont {X.}~\bibnamefont {Waintal}},\ }\bibfield  {title} {\bibinfo {title} {Numerical simulations of time-resolved quantum electronics},\ }\href {https://doi.org/10.1016j.physrep.2013.09.001} {\bibfield  {journal} {\bibinfo  {journal} {Phys. Rep.}\ }\textbf {\bibinfo {volume} {534}},\ \bibinfo {pages} {1} (\bibinfo {year} {2014})}\BibitemShut {NoStop}%
\bibitem [{\citenamefont {Sun}\ and\ \citenamefont {Linder}(2023)}]{Sun2023b}%
  \BibitemOpen
  \bibfield  {author} {\bibinfo {author} {\bibfnamefont {C.}~\bibnamefont {Sun}}\ and\ \bibinfo {author} {\bibfnamefont {J.}~\bibnamefont {Linder}},\ }\bibfield  {title} {\bibinfo {title} {Spin pumping from a ferromagnetic insulator into an altermagnet},\ }\href {https://doi.org/10.1103/PhysRevB.108.L140408} {\bibfield  {journal} {\bibinfo  {journal} {Phys. Rev. B}\ }\textbf {\bibinfo {volume} {108}},\ \bibinfo {pages} {L140408} (\bibinfo {year} {2023})}\BibitemShut {NoStop}%
\bibitem [{\citenamefont {Zhang}\ \emph {et~al.}(2024)\citenamefont {Zhang}, \citenamefont {Cui}, \citenamefont {Li}, \citenamefont {Duan}, \citenamefont {Li}, \citenamefont {Yu},\ and\ \citenamefont {Yao}}]{Zhang2024}%
  \BibitemOpen
  \bibfield  {author} {\bibinfo {author} {\bibfnamefont {R.-W.}\ \bibnamefont {Zhang}}, \bibinfo {author} {\bibfnamefont {C.}~\bibnamefont {Cui}}, \bibinfo {author} {\bibfnamefont {R.}~\bibnamefont {Li}}, \bibinfo {author} {\bibfnamefont {J.}~\bibnamefont {Duan}}, \bibinfo {author} {\bibfnamefont {L.}~\bibnamefont {Li}}, \bibinfo {author} {\bibfnamefont {Z.-M.}\ \bibnamefont {Yu}},\ and\ \bibinfo {author} {\bibfnamefont {Y.}~\bibnamefont {Yao}},\ }\bibfield  {title} {\bibinfo {title} {Predictable gate-field control of spin in altermagnets with spin-layer coupling},\ }\href {https://doi.org/10.1103/PhysRevLett.133.056401} {\bibfield  {journal} {\bibinfo  {journal} {Phys. Rev. Lett.}\ }\textbf {\bibinfo {volume} {133}},\ \bibinfo {pages} {056401} (\bibinfo {year} {2024})}\BibitemShut {NoStop}%
\bibitem [{\citenamefont {\ifmmode~\check{S}\else \v{S}\fi{}mejkal}\ \emph {et~al.}(2023)\citenamefont {\ifmmode~\check{S}\else \v{S}\fi{}mejkal}, \citenamefont {Marmodoro}, \citenamefont {Ahn}, \citenamefont {Gonz\'alez-Hern\'andez}, \citenamefont {Turek}, \citenamefont {Mankovsky}, \citenamefont {Ebert}, \citenamefont {D'Souza}, \citenamefont {\ifmmode~\check{S}\else \v{S}\fi{}ipr}, \citenamefont {Sinova},\ and\ \citenamefont {Jungwirth}}]{Smejkal2022c}%
  \BibitemOpen
  \bibfield  {author} {\bibinfo {author} {\bibfnamefont {L.}~\bibnamefont {\ifmmode~\check{S}\else \v{S}\fi{}mejkal}}, \bibinfo {author} {\bibfnamefont {A.}~\bibnamefont {Marmodoro}}, \bibinfo {author} {\bibfnamefont {K.-H.}\ \bibnamefont {Ahn}}, \bibinfo {author} {\bibfnamefont {R.}~\bibnamefont {Gonz\'alez-Hern\'andez}}, \bibinfo {author} {\bibfnamefont {I.}~\bibnamefont {Turek}}, \bibinfo {author} {\bibfnamefont {S.}~\bibnamefont {Mankovsky}}, \bibinfo {author} {\bibfnamefont {H.}~\bibnamefont {Ebert}}, \bibinfo {author} {\bibfnamefont {S.~W.}\ \bibnamefont {D'Souza}}, \bibinfo {author} {\bibfnamefont {O.}~\bibnamefont {\ifmmode~\check{S}\else \v{S}\fi{}ipr}}, \bibinfo {author} {\bibfnamefont {J.}~\bibnamefont {Sinova}},\ and\ \bibinfo {author} {\bibfnamefont {T.}~\bibnamefont {Jungwirth}},\ }\bibfield  {title} {\bibinfo {title} {Chiral magnons in altermagnetic {${\mathrm{RuO}}_{2}$}},\ }\href {https://doi.org/10.1103/PhysRevLett.131.256703} {\bibfield  {journal} {\bibinfo  {journal} {Phys. Rev. Lett.}\
  }\textbf {\bibinfo {volume} {131}},\ \bibinfo {pages} {256703} (\bibinfo {year} {2023})}\BibitemShut {NoStop}%
\bibitem [{\citenamefont {Jin}\ \emph {et~al.}()\citenamefont {Jin}, \citenamefont {Yang}, \citenamefont {Zeng}, \citenamefont {Cao},\ and\ \citenamefont {Yan}}]{Jin2023}%
  \BibitemOpen
  \bibfield  {author} {\bibinfo {author} {\bibfnamefont {Z.}~\bibnamefont {Jin}}, \bibinfo {author} {\bibfnamefont {H.}~\bibnamefont {Yang}}, \bibinfo {author} {\bibfnamefont {Z.}~\bibnamefont {Zeng}}, \bibinfo {author} {\bibfnamefont {Y.}~\bibnamefont {Cao}},\ and\ \bibinfo {author} {\bibfnamefont {P.}~\bibnamefont {Yan}},\ }\href@noop {} {\bibinfo {title} {Cavity-induced strong magnon-magnon coupling in altermagnets}},\ \Eprint {https://arxiv.org/abs/2307.00909 (2023)} {arXiv:2307.00909 (2023)} \BibitemShut {NoStop}%
\bibitem [{\citenamefont {Papaj}(2023)}]{Papaj2023}%
  \BibitemOpen
  \bibfield  {author} {\bibinfo {author} {\bibfnamefont {M.}~\bibnamefont {Papaj}},\ }\bibfield  {title} {\bibinfo {title} {Andreev reflection at the altermagnet-superconductor interface},\ }\href {https://doi.org/10.1103/PhysRevB.108.L060508} {\bibfield  {journal} {\bibinfo  {journal} {Phys. Rev. B}\ }\textbf {\bibinfo {volume} {108}},\ \bibinfo {pages} {L060508} (\bibinfo {year} {2023})}\BibitemShut {NoStop}%
\bibitem [{\citenamefont {Sun}\ \emph {et~al.}(2023)\citenamefont {Sun}, \citenamefont {Brataas},\ and\ \citenamefont {Linder}}]{Sun2023a}%
  \BibitemOpen
  \bibfield  {author} {\bibinfo {author} {\bibfnamefont {C.}~\bibnamefont {Sun}}, \bibinfo {author} {\bibfnamefont {A.}~\bibnamefont {Brataas}},\ and\ \bibinfo {author} {\bibfnamefont {J.}~\bibnamefont {Linder}},\ }\bibfield  {title} {\bibinfo {title} {Andreev reflection in altermagnets},\ }\href {https://doi.org/10.1103/PhysRevB.108.054511} {\bibfield  {journal} {\bibinfo  {journal} {Phys. Rev. B}\ }\textbf {\bibinfo {volume} {108}},\ \bibinfo {pages} {054511} (\bibinfo {year} {2023})}\BibitemShut {NoStop}%
\bibitem [{\citenamefont {Troncoso}\ \emph {et~al.}(2019)\citenamefont {Troncoso}, \citenamefont {Brataas},\ and\ \citenamefont {Duine}}]{Troncoso2019}%
  \BibitemOpen
  \bibfield  {author} {\bibinfo {author} {\bibfnamefont {R.~E.}\ \bibnamefont {Troncoso}}, \bibinfo {author} {\bibfnamefont {A.}~\bibnamefont {Brataas}},\ and\ \bibinfo {author} {\bibfnamefont {R.~A.}\ \bibnamefont {Duine}},\ }\bibfield  {title} {\bibinfo {title} {Many-body theory of spin-current driven instabilities in magnetic insulators},\ }\href {https://doi.org/10.1103/PhysRevB.99.104426} {\bibfield  {journal} {\bibinfo  {journal} {Phys. Rev. B}\ }\textbf {\bibinfo {volume} {99}},\ \bibinfo {pages} {104426} (\bibinfo {year} {2019})}\BibitemShut {NoStop}%
\bibitem [{\citenamefont {Rohling}\ \emph {et~al.}(2018)\citenamefont {Rohling}, \citenamefont {Fj\ae{}rbu},\ and\ \citenamefont {Brataas}}]{Rohling2018}%
  \BibitemOpen
  \bibfield  {author} {\bibinfo {author} {\bibfnamefont {N.}~\bibnamefont {Rohling}}, \bibinfo {author} {\bibfnamefont {E.~L.}\ \bibnamefont {Fj\ae{}rbu}},\ and\ \bibinfo {author} {\bibfnamefont {A.}~\bibnamefont {Brataas}},\ }\bibfield  {title} {\bibinfo {title} {Superconductivity induced by interfacial coupling to magnons},\ }\href {https://doi.org/10.1103/PhysRevB.97.115401} {\bibfield  {journal} {\bibinfo  {journal} {Phys. Rev. B}\ }\textbf {\bibinfo {volume} {97}},\ \bibinfo {pages} {115401} (\bibinfo {year} {2018})}\BibitemShut {NoStop}%
\bibitem [{\citenamefont {Griffiths}(2011)}]{Griffiths2011}%
  \BibitemOpen
  \bibfield  {author} {\bibinfo {author} {\bibfnamefont {D.~J.}\ \bibnamefont {Griffiths}},\ }\bibfield  {title} {\bibinfo {title} {Dynamic dipoles},\ }\href {https://doi.org/10.1119/1.3591336} {\bibfield  {journal} {\bibinfo  {journal} {Am. J. Phys.}\ }\textbf {\bibinfo {volume} {79}},\ \bibinfo {pages} {867} (\bibinfo {year} {2011})}\BibitemShut {NoStop}%
\bibitem [{\citenamefont {Sambles}\ \emph {et~al.}(1982)\citenamefont {Sambles}, \citenamefont {Elsom},\ and\ \citenamefont {Jarvis}}]{Sambles1982}%
  \BibitemOpen
  \bibfield  {author} {\bibinfo {author} {\bibfnamefont {J.~R.}\ \bibnamefont {Sambles}}, \bibinfo {author} {\bibfnamefont {K.~C.}\ \bibnamefont {Elsom}},\ and\ \bibinfo {author} {\bibfnamefont {D.~J.}\ \bibnamefont {Jarvis}},\ }\href {https://doi.org/10.1098/rsta.1982.0016} {\bibfield  {journal} {\bibinfo  {journal} {Philos. Trans. A: Math. Phys. Eng. Sci.}\ }\textbf {\bibinfo {volume} {304}},\ \bibinfo {pages} {365} (\bibinfo {year} {1982})}\BibitemShut {NoStop}%
\end{thebibliography}%

\end{document}